\documentclass[sigconf]{acmart}

\usepackage{graphicx}
\usepackage{balance}
\AtBeginDocument{%
  \providecommand\BibTeX{{%
    \normalfont B\kern-0.5em{\scshape i\kern-0.25em b}\kern-0.8em\TeX}}}

\copyrightyear{2020}
\acmYear{2020}
\setcopyright{acmcopyright}\acmConference[CPSS '20]{Proceedings of the 6th ACM Cyber-Physical System Security Workshop}{October 6, 2020}{Taipei, Taiwan}
\acmBooktitle{Proceedings of the 6th ACM Cyber-Physical System Security Workshop (CPSS '20), October 6, 2020, Taipei, Taiwan}
\acmPrice{15.00}
\acmDOI{10.1145/3384941.3409587}
\acmISBN{978-1-4503-7608-2/20/10}

\begin{document}
\fancyhead{}
\title[Evaluating Cascading Impact of Attacks on Resilience of ICS]{Evaluating Cascading Impact of Attacks on Resilience of Industrial Control Systems: A Design-Centric Modeling Approach}

\author{Zhongyuan Hau}
\affiliation{%
  \institution{Imperial College London, United Kingdom}
}\email{zy.hau17@imperial.ac.uk}

\author{John Henry Castellanos}
\affiliation{%
  \institution{Singapore University of Technology and Design, Singapore}
}
\email{john\_castellanos@mymail.sutd.edu.sg}

\author{Jianying Zhou}
\affiliation{%
  \institution{Singapore University of Technology and Design, Singapore}
}
\email{jianying\_zhou@sutd.edu.sg}

\renewcommand{\shortauthors}{Z. Hau, et al.}

\begin{abstract}
A design-centric modeling approach was proposed to model the behaviour of the physical processes controlled by Industrial Control Systems (ICS) and study the cascading impact of data-oriented attacks. A threat model was used as input to guide the construction of the CPS model where control components which are within the adversary's intent and capabilities are extracted. The relevant control components are subsequently modeled together with their control dependencies and operational design specifications. The approach was demonstrated and validated on a water treatment testbed. Attacks were simulated on the testbed model where its resilience to attacks was evaluated using proposed metrics such as Impact Ratio and Time-to-Critical-State. From the analysis of the attacks, design strengths and weaknesses were identified and design improvements were recommended to increase the testbed's resilience to attacks.
\end{abstract}



\maketitle

\section{Introduction}
Cyber-physical systems (CPS) are systems that comprise of networked computational and communication devices that interacts with the physical environment they are deployed in. Industrial Control Systems (ICS) belongs to the family of CPS, and they commonly comprise of controllers, sensors and actuators which are used in a variety of industries to facilitate the automated monitoring and control of physical processes. With advancement in technology, there is increased demand for the modernization of these systems to improve efficiency through enhanced connectivity, monitoring and control capabilities \cite{krotofil2014you}. However, enhanced connectivity also increases the attack surface of these systems, making them more vulnerable to cyber attacks. Cyber attacks on CPS would not only result in the loss of confidentiality and integrity of information, it could have a far worse consequence of damaging the physical environment and inflicting harm on human lives \cite{cardenas2011attacks,krotofil2013resilience,krotofil2015process}.

Acknowledging the need to address security of CPS, there has been extensive research on securing CPS from different aspects such as hardware security, network-level intrusions detection and more recently, process-data based attack detection \cite{stouffer2011guide,giraldo2018survey, hau2019exploiting, ahmed2018noise}. While securing CPS is important, in the unfortunate event that an attacker has successfully bypass defense systems, it is also critical to evaluate the impact of disruption from the attack and formulate counter-measures and recovery strategies. For an attacker to achieve their desired disruptive results, it requires the understanding of the workings of the underlying systems that make up the CPS, such as the control logic and process behaviours \cite{orojloo2017method}. Attackers that target CPS often interact with the system and exploit failure modes that disrupts the physical processes. As such, there is a need for a model to study how the attacker's interaction with the CPS would impact the physical processes. 

Physical processes controlled by CPS are complex with numerous sub-processes that may span across vast geographical areas. the sub-processes are inter-connected either physically or by the control dependencies from the control logic of the controllers \cite{10.1145/3274694.3274745}. As such, attacks on any individual component or subset of components, would result in a cascading effect within the system. CPS are safety-critical systems and an important aspect of security research is to study the impact and consequences of attacks on the physical environment \cite{krotofil2015process,orojloo2017method}. With a model that aids the understanding of the interaction of adversary with the physical system and demonstrate the propagation effect of cyber attacks, it can help in the design of defences and counter-measures that improves the system's overall resilience.

In all, while most research focus on securing CPS from attacks, there is an equally important need to understand and quantify the impact of attacks in CPS. To this end, this paper presents a design-centric approach to construct a model that simulates the process dynamics and propagation of attacks across the sub-processes of a process system. \textbf{In specific, we are studying the propagation of effects of integrity attack on system state across the sub-systems of a CPS (i.e the impact of a compromised sensor value used in the control-action decision of multiple sub-systems).} A quantitative measure of resilience to cyber attacks (\textbf{Impact Ratio)} was introduced in the paper, this metric is similar to those used in studies of networked infrastructure \cite{ouyang2014multi,reed2009methodology,o2007critical}, the ratio indicates the deviation of system operation from the norm within a period of time. We also quantified the ability of the system to withstand attacks before an undesirable consequence occurs by measuring the \textbf{time-to-critical-state} whilst under a sustained cyber attack \cite{castellanos2019AModular}.

The main contributions of this paper are as follows:
\begin{enumerate}
    \item A modeling approach that takes into consideration a threat model to guide the construction of a CPS model. The model is subsequently used for simulations to evaluate the cascading impact of attacks in a CPS. The design driven modeling approach  combines the system's control strategy and operating specifications to model the process dynamics.
    \item A systematic approach to evaluate the impact of attacks and measure the resilience of the system under various attack scenarios.
\end{enumerate}
We applied the proposed modeling approach on the Secure Water Treatment Testbed (SWaT) \cite{SWaT} to validate the modeling approach. Subsequently, we attempted to quantify the impact of attacks by providing a measurement of the system's resilience to the disruptions under the various attack scenarios. 

The organization of the subsequent sections of the paper are as follows. Section \ref{sec:related_work} reviews some of the related work. In Section \ref{sec:approach}, we present the modeling approach and evaluation methodology. In Section \ref{sec: Validate} , the modeling approach was implemented for a testbed and validated against historical data. In Section \ref{sec:SA}, various attacks were conducted and sensitivity analysis were performed to evaluate their impacts on system resilience. Finally in Section \ref{sec:conclusion} we provide concluding remarks and discuss some future work.

\section{Related Work}\label{sec:related_work}
In this section, we review some of the related work on approaches undertaken to measure and analyse impact of cyber attacks on CPS. 

Cardenas \textit{et al.} in \cite{cardenas2011attacks} demonstrated the use of process models to study the interaction between control system and the physical processes. The model of a single stage chemical reactor process was used to determine sensors to attack in order to drive the system to an unsafe state. The approach taken was to perform integrity and denial of service attacks on all sensors to evaluate each attack's ability to drive the system to an unsafe state. Krotofil and Cardenas in \cite{krotofil2013resilience} also investigated the impact of attacks on control systems using process models. Attacks were simulated on a single-stage chemical reactor process and empirical analysis on the effects of attacks on physical process behaviour were conducted. These works, however, do not discuss the coupling effects of attacks that results from control loop dependencies.

In \cite{genge2015system}, Genge \textit{et al.} proposed a methodology of assessing the impact of attacks by measuring the cross-covariances of control variables before and after the system is perturbed. While this method provides insights on how the impact of attack propagates through the system via the relationship between control variables, it does not translate to the consequence of attacks on system performance.

 Milosevic \textit{et al.} in \cite{milovsevic2018quantifying} proposed a framework to measure impact of attack  on stochastic linear control systems using the infinity norm of critical states over a time window. The impact is measured by how much the critical states of the system deviates from the steady state over a period of time steps while remaining undetected by an anomaly detector. The impact metric provides information of the extent to which the system is perturbed during an attack but does not give resolution on the impact on the physical process.

In \cite{orojloo2017method}, Orojloo and Azgomi proposed a modeling approach that considers the systems dynamics and control dependencies between the various components in the cyber-physical systems. The model built was used to perform sensitivity analysis to understand the system behaviour under various attack scenarios, providing insights to vulnerable control loops. The impact on system's physical parameters by attacks on specific components on the system was subsequently used to evaluate the component's criticality for successful attacks.

Adepu and Mathur in \cite{adepu2016investigation} studied the responses of a water treatment plant to various single point attacks. The effects of attack propagation in terms of the number of components in the system affected were analyzed. System behaviour such as changes in physical process metrics during attacks were investigated. The results of the study were used to propose attack detection mechanisms that were based on physical properties of the system.

In \cite{Ahmed:2017:MAD:3052973.3053011}, Ahmed \textit{et al.} used Linear Time Invariant (LTI) stochastic difference equations to obtain a set of discrete time state space model that describes the behaviour of water level in the tanks of their simulated water distribution network. However, this modeling does not take into account the control strategy that decides on the control action. While the LTI state space modeling provides good estimation of the system state, there is a need to incorporate modeling of the control strategy which could be exploited by attackers.

Comparing the works that focus on modeling and measuring impact of attacks on cyber-physical systems, we identified that there is a gap in providing an approach to quantitatively assess cascading impact of attacks across the multiple processes within a system and provide measurements of resilience of the system to attacks with respect to safety and performance. 

\vspace{-0.2cm}
\section{Approach}\label{sec:approach}
In this section, we outline the approach taken to build a threat intent guided model of a cyber-physical system which incorporates the interaction of process's control strategy with the physical process and the attacker's capabilities. This allows us to conduct simulations of attacks on the physical process via the control system to analyse how the impact of attacks propagates in the system and study the system's resilience to cyber attacks. 

\subsection{Threat Modeling}
Threat modeling starts by describing the intention and capabilities of the attacker and the output would be a subset of physical components and related control strategies that the adversary would be able to attack and affect.

\subsubsection{Definition 1.}We define the intention of the attacker as adversarial goals that the attacker wants to achieve from an attack. We assume that the attacker's intention is to disrupt the operational capability of the system, and the disruption can be measured by sensors in the system. In the first step, the set of sensors that measure the affected operational metrics identified. 

The set of $n$ sensors $S_{metric}$ that measure the impact of attack  directly related to the attacker's goal is a subset of all the sensors $S$ in the system.
\begin{equation}
    S_{metric} = \{ s_{1}, s_{2}, \dots , s_{n} \}, \forall i =1,..,n \land s_{i} \in S
\end{equation}

The second step would be to find control statements that  relates the affected the operational metrics measured by $S_{metric,i}$ to other components in the system. With the knowledge of the design of the control strategy for normal operations, we are able to extract Control Statements ($CS$) that relates $S_{metric, i}$ to other components in the system.

\begin{equation}
    CS_{j} = \; <S_{metric, i}, S_{CS,j} , A_{CS,j}, Cond > , \forall j = 1,..., m
\end{equation}
where :
\begin{list}{}
    \item $m$ is the total number of control statements
    \item $S_{metric, i}$ is the $i$th metric in $S_{metric}$
    \item $S_{CS, j} \subset S$ , and $S$ is the set of all Sensors in the system
    \item $A_{CS, j} \subset A$ , and $A$ is the set of all Actuators in the system
    \item $ Cond $ are conditions that relates states of $S_{CS}$ and $s_{metric, i}$ to changes to state of $A_{CS}$
\end{list}

\subsubsection{Definition 2.}We define the capability ($Cap$) of the attacker as a subset of sensor and actuators that the attacker has control over, meaning that the attacker is able to manipulate the data received and sent by the component. 

\begin{equation}
    Cap = \; < S_{attacker} , A_{attacker} >
\end{equation}
where :
\begin{list}{}
    \item $S_{attacker} \subseteq S$ , and $S$ is the set of all Sensors in the system
    \item $A_{attacker} \subseteq A$ , and $A$ is the set of all Actuators in the system
\end{list}

\subsubsection{Output for model.} From the adversary's threat intent and capabilities, we are able to obtain set of sensors, actuators and control statements that are relevant for modeling the interaction of the attacker with the process dynamics and control loops :

\begin{list}{}
    \item $S_{model} =  S_{CS,j}  \cap S_{attacker} , \forall j = 1,\dots,m $ 
    \item $A_{model} = A_{CS, j} \cap A_{attacker} , \forall j = 1,\dots,m $
    \item $CS_{model} = \; <s_{x}, a, cond > \cup <s, a_{x}, cond >$
    \begin{list}{}
        \item $\forall s_{x} \in S_{model} , \forall  a_{x} \in A_{model}  \land a \in A \land s \in S $
    \end{list}
\end{list}

These are the sets of sensors and actuators in the CPS that controls the various sub-processes which the attacker has the ability to influence due to the control and physical dependencies of the components. These components and their physical processes are of interest to us for modeling so that we can evaluate the impact of the attacker's interaction with the CPS.

\subsection{Modeling Process System Dynamics}
The approach to construct the model of a process system draws inspiration from System Dynamics, an area of research that aims to study and describe the behaviour of complex systems over time. With fundamentals of control theory  and theory of non-linear dynamical systems as the foundation of System Dynamics \cite{sterman2001system}, it is a suitable modeling approach for studying interaction of attacker's input on process dynamics.

System Dynamics models actual flow of physical entities or information, regulated by feedback loops \cite{forrester1997industrial}. Tools used to describe the system are Stock and Flow Diagram (SFD) and Causal Loop Diagrams (CLD). The relationship between system variables is represented by a CLD, where qualitative analysis of the dynamics can be performed. The flow and accumulation characteristics can be studied using a SFD where quantitative analysis can be performed using simulations.  From the output of threat modeling phase, we have a set of Control Statements which relates sensors and actuators. Sensors in a dynamical systems measures the stock and flow, whereas the actuators changes these measurements in the system. In addition, we utilize the inputs from design specifications of the process system to construct the system model with information of how components are physically related such as physical connection and operating parameters that affect stock and flow.

\vspace{-0.35cm}
\begin{figure}[htbp]
\centering
\subfigure[Tank Example]{\label{fig:tank_example}\includegraphics[scale= 0.39]{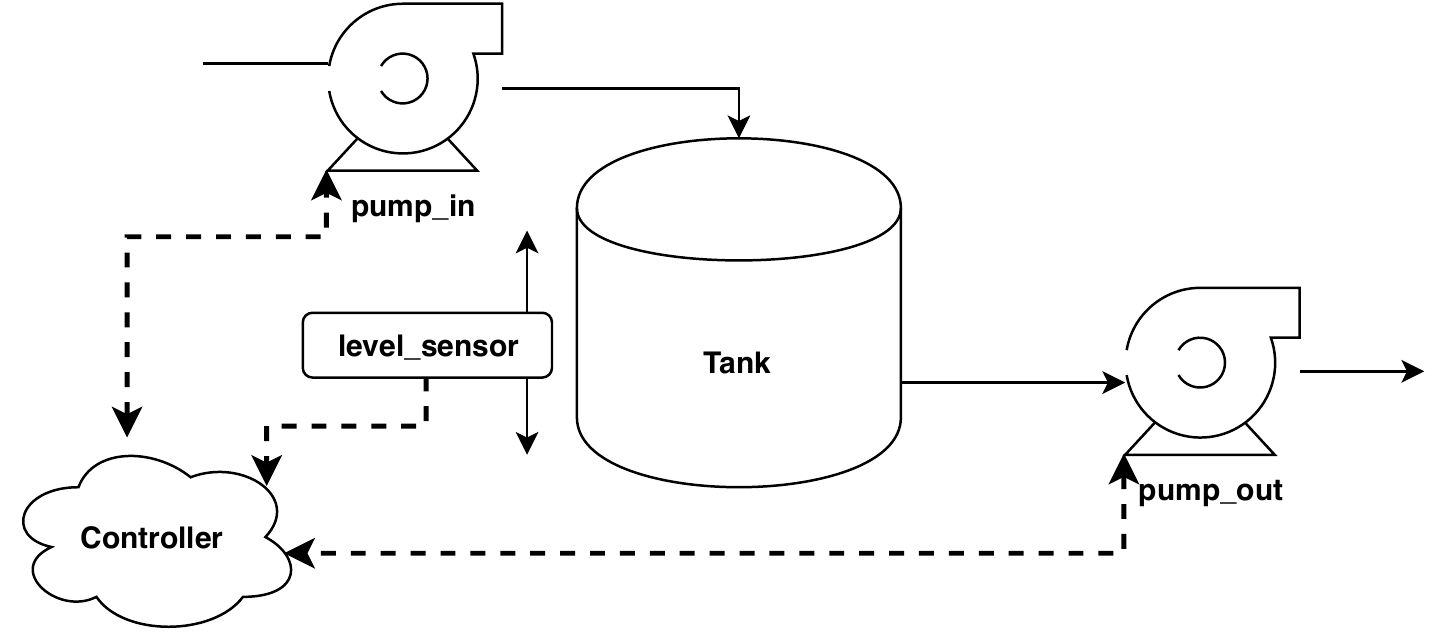}}
\subfigure[Tank example with system state representation in "cyber realm" under attack]{\label{fig:tank_example_attack}\includegraphics[scale = 0.39]{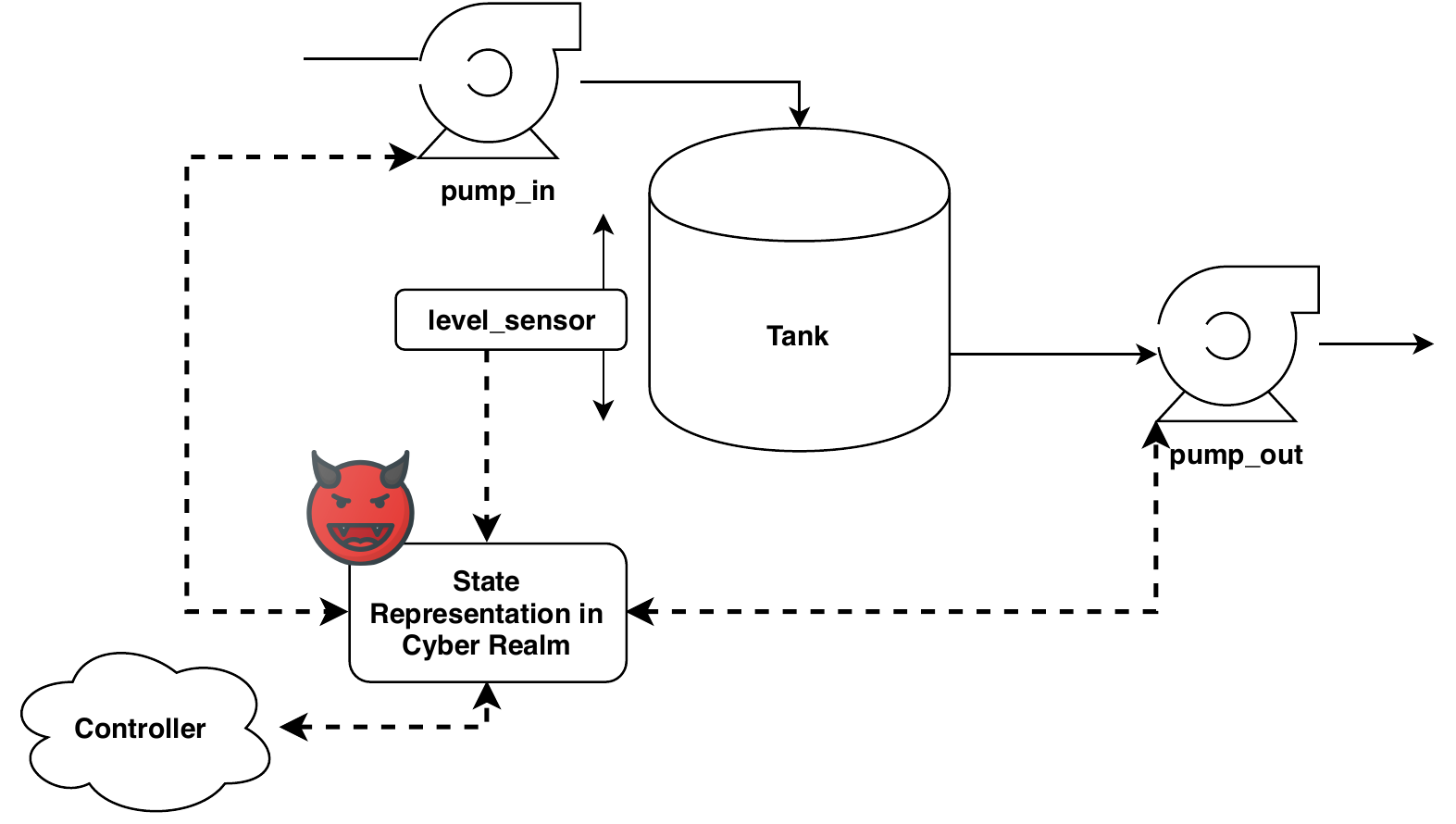}}
\vspace{-0.45cm}
\caption{Illustrative Single Tank Example}
\label{fig:metric_exp2_p1}
\end{figure}
\vspace{-0.35cm}

We use a simple tank example to illustrate the modeling approach. In Figure \ref{fig:tank_example}, we have a system comprising of a tank with an inlet pump and an outlet pump, the level of liquid in the tank is measured by a level sensor. The rate of inflow and outflow determines the rate of tank level increase and decrease; these are described by design specifications of these components.

\begin{minipage}[htbp]
{0.9\linewidth}
\begin{lstlisting}[language = python, caption = Tank Level Control Strategy, label = code] 
# Control Statement 1 
# tank level is high, stop inflow and start outflow
if level_sensor.get() >= level_high : 
    pump_in.off()
    pump_out.on()

# Control Statement 2
# tank level is low, start inflow and stop outflow
if level_sensor.get() <= level_low : 
    pump_in.on()
    pump_out.off()
\end{lstlisting}
\end{minipage}

Control statements (Listing \ref{code}) exists such that the tank level does not rise or fall below pre-defined levels. This results in causal loops that serves to regulate the liquid level in the tank. The SFD, together with design specifications of the system, were used to build a simulation model for quantitative analysis.

\subsubsection{Modeling Adversarial Interactions and Physical Impact.} Under normal operation, when there is absence of attacks and faulty components, the state of the system perceived by the controller is approximately identical to the actual physical state of the process. However, during adversarial cyber attacks, the perceived states of the system can be manipulated to disrupt the process dynamics by exploiting control loops to cause the system to react to the perceived state.

During attack, the perceived state of the system by the controller deviates from the actual physical state and would not be able to detect actual changes to the physical state caused by control instructions. In order to study the physical impact of the attacks, there is a need to model both the cyber and physical representations of the system. Under normal conditions, the cyber representation is a close approximation of the physical representation. When under cyber attack, the physical representation is the actual physical state whereas the cyber representation is controlled by the attacker.

Using the same tank example, in Figure \ref{fig:tank_example_attack}, we duplicate the states of every component. The level sensor measures the actual physical level of the tank whereas the cyber representation of the sensor is manipulated by the attacker. The controller perceives the "cyber realm" system state and give instructions to actuators that act on the physical process. With both cyber and physical representations of the system, we are able to study the impact of attacks on physical process with system state manipulation in cyber realm. 

In all, the modeling of the system dynamics provides the prediction of the system state, taking into account previous system and control action, which is similar to the modeling using Linear Time Invariant state space modeling. The modeling approach that encodes control dependencies of system states and control strategies provides an additional information of predicted control action taken by the controllers based on current system state as the control dependencies and strategies are encoded into the model. This allows us to build a model that is a "digital twin" of the physical CPS we are modeling with respect to the pre-defined threat capabilities and goal.

\subsubsection{ Toy Example With Single Tank System.} 
\begin{figure}[htbp]
\centering
\includegraphics[scale=0.6]{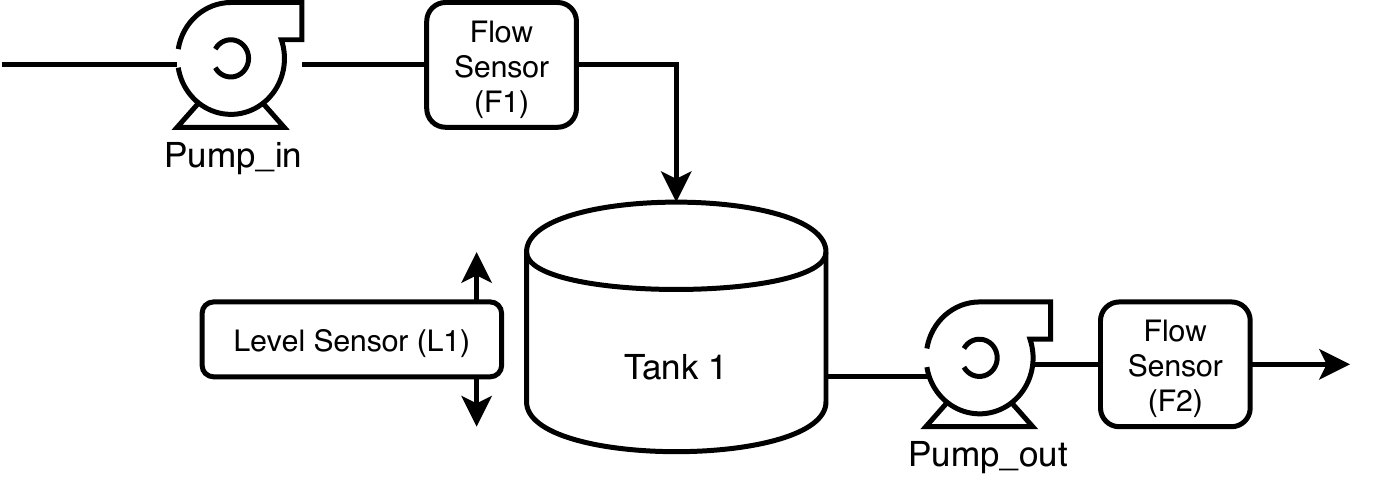}
\caption{Illustrative Single Tank Example}
\label{fig:single_tank}
\end{figure}

Using the single tank system as an example with additional flow sensors after each pump (See Figure \ref{fig:single_tank}), we describe the process of obtaining the set of sensors and actuators to model. We assume that the attacker's goal is to disrupt the throughput of the Single Tank System and the attacker is capable of manipulating the level sensor L1. The first step would be to identify the operational metric that is relevant to the attacker's goal, which is the flow sensor F2 that measures the flow rate of water out of the system. F2 is determined by the operation of Pump\_out. Pump\_out 's operation is governed by Control Statement 2 in Listing 1. Pump\_in's operation is also dependent on the level sensor (Control Statement 1 in Listing 1) and the flow sensor F1's reading is dependent on the operation of Pump\_in. 

As such, in this simple toy example, the set of sensors are all the sensors L1, F1 and F2, the set of actuators are both pumps and the control statements are the 2 control statements in Listing 1. In a larger and more complex CPS, there would be sub-systems that are not 'reachable' by the attacker and hence would not be modelled. 

If we were to model the process dynamics of the system under attack, the attacker would spoof the measurement of L1 to be low so that Pump\_out is off. However, this would result in Pump\_in to remain in the state 'On', which would result in the overflow of the tank. In the next section, we discuss how we quantify the impact of attacks and characterise the resilience of systems.

\subsection{Attack Impact Measurement}
With the model, we are able to simulate the process dynamics under both scenarios of normal conditions and when the system is under attack. To help us study the interaction of the attacker with the system behaviour and understand the resilience of the system to attacks, it would be important for us to be able to measure the impact of attack in comparison to normal conditions. Resilience with respect to cyber-attacks is defined as the ability of the system to absorb, recover and adapt to an attack that causes degradation in performance \cite{kott2019cyber}. In this section, we discuss two methods of quantifying impact of attacks. 

\subsubsection{Area Between Operational Curves.} 
\begin{figure}[htbp]
\centering
\includegraphics[width = 0.5\textwidth]{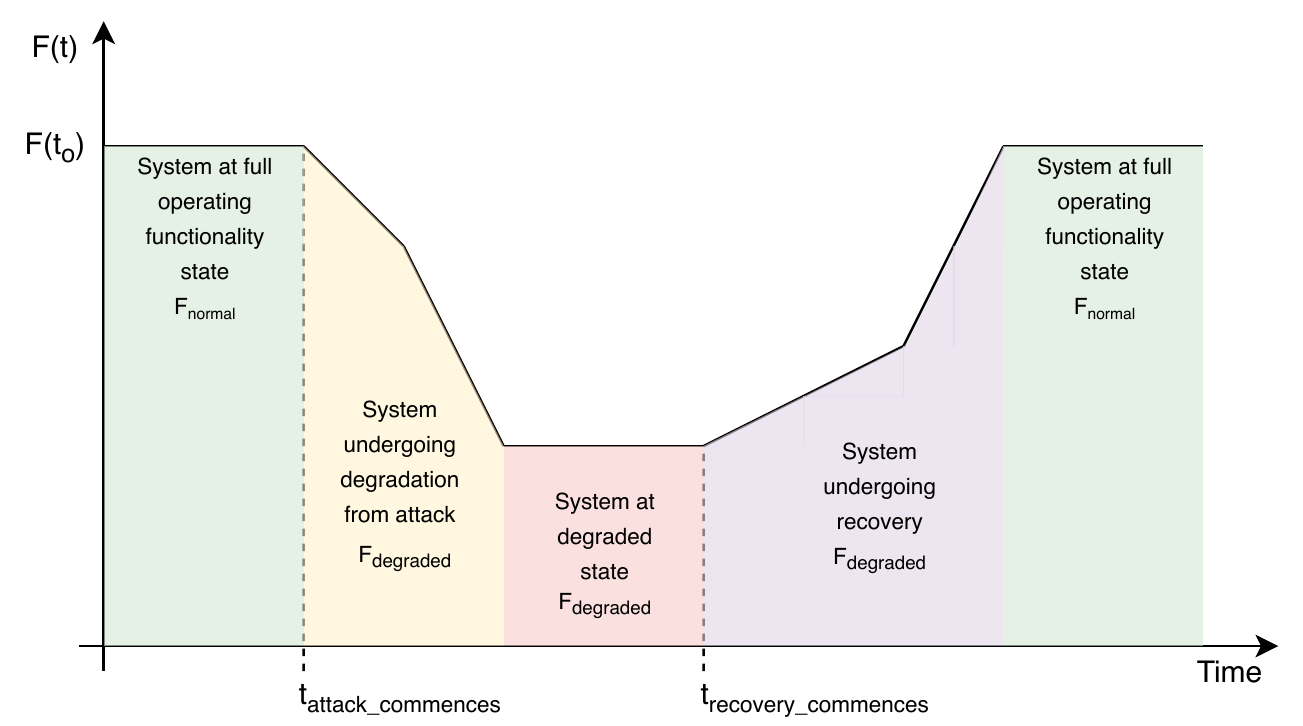}
\caption{Operational Level F(t), Transition Over Time.}
\label{fig:resilience_metric1}
\end{figure}

In many recent studies on resilience of networked infrastructure such as power systems and Information Technology systems, resilience is quantified by a time-dependent metric of changes in operational levels over time \cite{reed2009methodology,ouyang2014multi}. Resilience is defined as the ability of the system to withstand, absorb, adapt to and recover from disturbances to the system \cite{o2007critical}. The resilience trapezoid (Figure \ref{fig:resilience_metric1}), shows how the operational level changes over time as it transits through the various phases from original state to system undergoing disruption, disrupted state, recovery state and recovered state.

In our work, since we are able to simulate system performance under both scenarios of normal and attack conditions, we are able to quantify the impact of the attack by measuring the divergence of the two curves. This could be measured using the difference in cumulative area of the normal operating curve and operating curve when the system is under attack. The operational levels used to calculate area under graph would be sensor measurements of the actual physical process (i.e. before duplication of state for cyber representation which can be manipulated by attacker). The difference in cumulative area will then be normalized against the area under the normal operating curves, this results in a value we define as \textbf{Impact Ratio}. 

\begin{equation}
    Impact \; Ratio = \frac{\int_{t_{start}}^{t_{end}} F_{degraded}(t) \: dt - \int_{t_{start}}^{t_{end}} F_{normal}(t) \: dt}{\int_{t_{start}}^{t_{end}} F_{normal}(t) dt}
\end{equation}
where :
\begin{list}{}
    \item $F_{normal}(t)$ is the operating curve (time-series of operational metric) under normal operating conditions
    \item $F_{degraded}(t)$ is the operating curve when system is at degraded state (Under disruption or during recovery)
    \item $t_{start}$ and $t_{end}$ are start and end times of analysis respectively
\end{list}

In the absence of attacks, there will be no deviation and impact ratio would be 0, whereas during an attack, the deviation in the two operating curves can be measured by the magnitude of change in the ratio. A negative ratio indicates decrease in the measured operational variable whilst under attack, conversely, a positive ratio indicates an increase in the measured operational variable.

\subsubsection{Time -  to - critical - state.}
 Critical states are states where a system's operational performance and/or safety has reached an unacceptable level. 
 We can 'measure' how fast a system can reach the nearest critical state by computing the time-to-critical-state~\cite{castellanos2019AModular} of the system from a particular time. A longer time-to-critical-state during an attack would indicate that the system has better capabilities to withstand the particular system disruption and hence, more resilient to the particular attack.

\section{Model Validation} \label{sec: Validate}
The approach outlined in Section \ref{sec:approach} was applied to model the Secure Water Treatment Testbed (SWaT) \cite{SWaT} using an attacker model and the operational specifications of the testbed as inputs. The attacker is assumed to have control of all the sensors and actuators in the testbed and the attacker's intention is to disrupt the flow of water through the testbed such that the \textit{\textbf{throughput of treated water from the water treatment plant is affected}}. As such, the attacker is able to perform sensor measurement and actuator command spoofing across the whole system to perform complex attacks that maximises damage. Consequentially, the impact on safety that arises from attacks such that the attacker's intention are fulfilled would be overflowing of tanks and dry running of centrifugal pumps.

\subsection{SWaT Testbed}
\begin{figure}[htbp]
\centering
\includegraphics[scale=0.2]{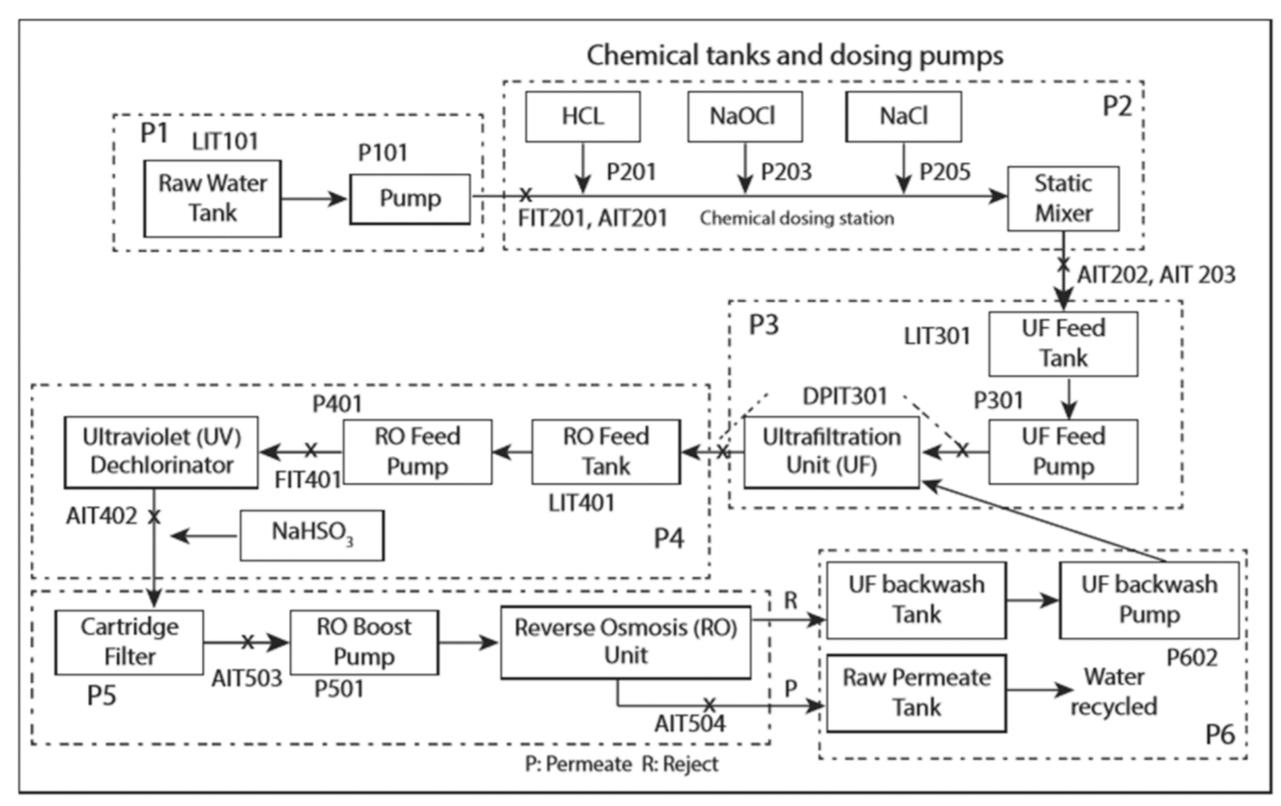}
\caption{Overview of Processes in SWaT. Figure from \cite{SWaT}}
\label{fig:swat_process}
\end{figure}

SWaT is a fully-functional six stage water treatment plant testbed with a throughput of 5 gallons/min (1.14 $m^3/hr$) of treated water that is primarily used by researchers to study attacks on ICS infrastructure. Figure \ref{fig:swat_process} provides an overview of the processes within the water treatment plant and the various components within each process (LIT : Level Indicator Transmitter, AIT : Analytical Indicator Transmitter, FIT : Flow Indicator Transmitter, P = Pump).

\subsection{Model Validation Approach}
In \cite{SWaT}, the work provides historical dataset of the testbed's state under normal operating conditions and under various attack scenarios. We utilize the dataset to validate our model, our model uses the same initial conditions as provided by the dataset and run the simulation for a stipulated time period. The experiments to validate are as follows :
\begin{enumerate}
    \item Normal Operating Conditions for 4 hours
    \item Attack 7: Single Stage, Single Point Attack (SSSP) on Process 3's LIT-301
    \item Attack 30: Multi-Stage, Multi-Point Attack (MSMP) on components in Process 1 and Process 2
\end{enumerate}
We have only simulated Process 1 to Process 5 as Process 6 serves as a recycle stage and out of scope of the study of impact on plant operations related to attacker's intent and capabilities.

As our goal is to investigate the effects of cascading impact of attacks that exploit the inter-dependencies of sensor values in control system, we have specifically chosen Attack 7 that spoofs a single sensor value that is used in control decisions for multiple stages in the plant. We have also chosen Attack 30 to demonstrate the ability of the model to simulate complex attacks that involves manipulation of both sensor and actuator values across multiple stages.

\subsubsection{Normal Operations.}
\begin{figure}[htbp]
\centering
\subfigure[FIT-101]{\label{fig:flow_4hrs_a}\includegraphics[scale = 0.46]{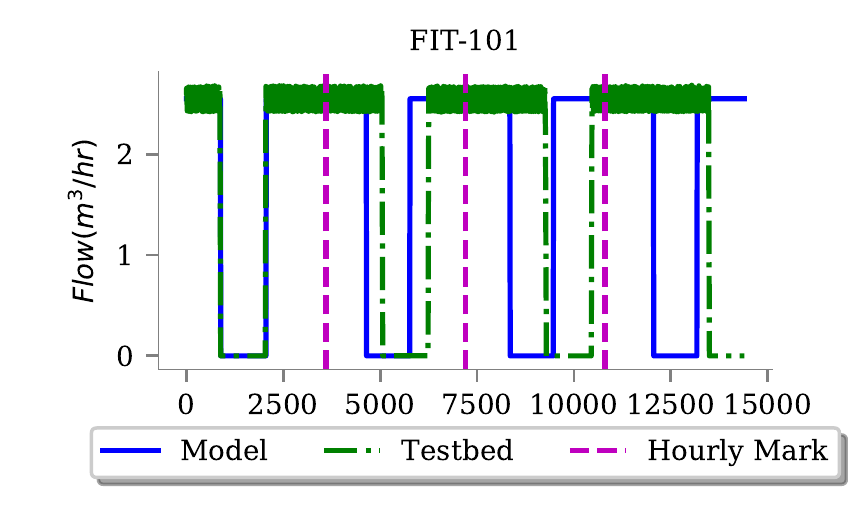}}
\subfigure[FIT-201]{\label{fig:flow_4hrs_b}\includegraphics[scale = 0.46]{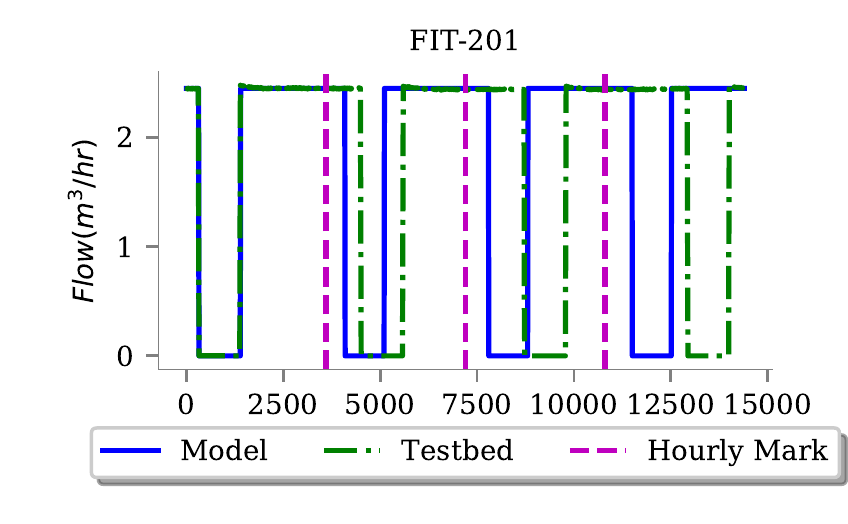}}
\subfigure[FIT-301]{\label{fig:flow_4hrs_c}\includegraphics[scale = 0.46]{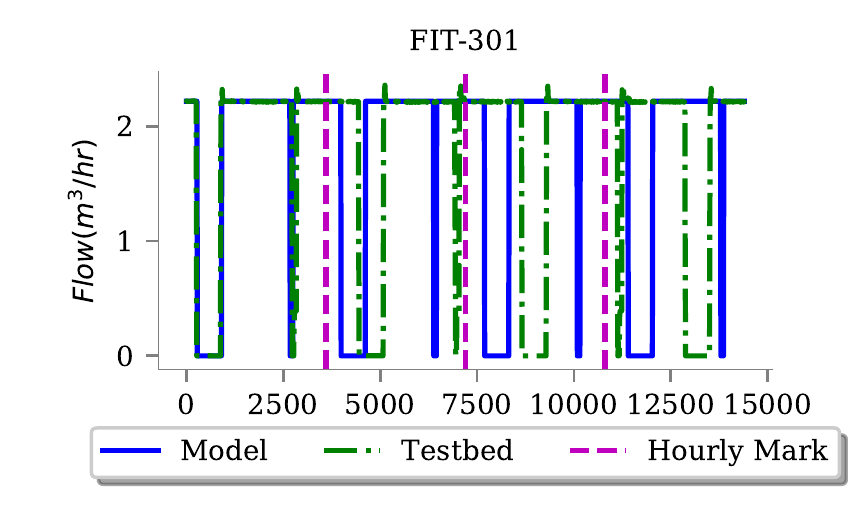}}
\subfigure[FIT-401]{\label{fig:flow_4hrs_d}\includegraphics[scale = 0.46]{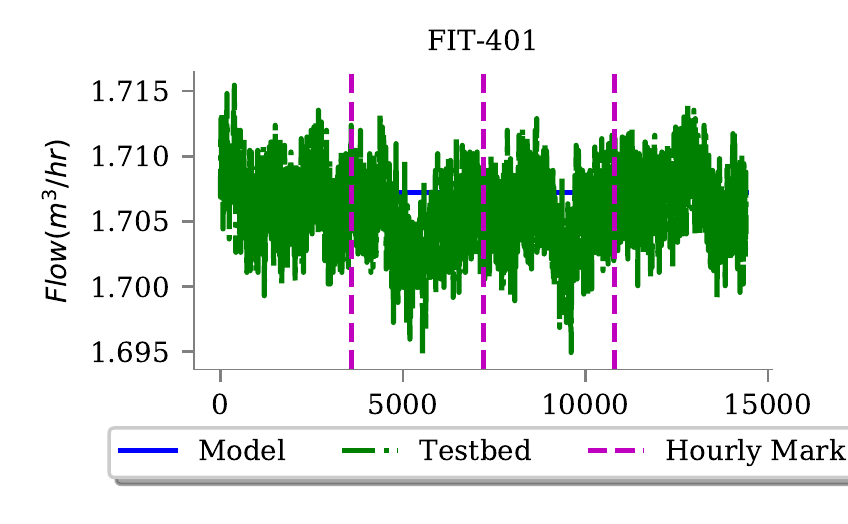}}
\subfigure[FIT-501]{\label{fig:flow_4hrs_e}\includegraphics[scale = 0.46]{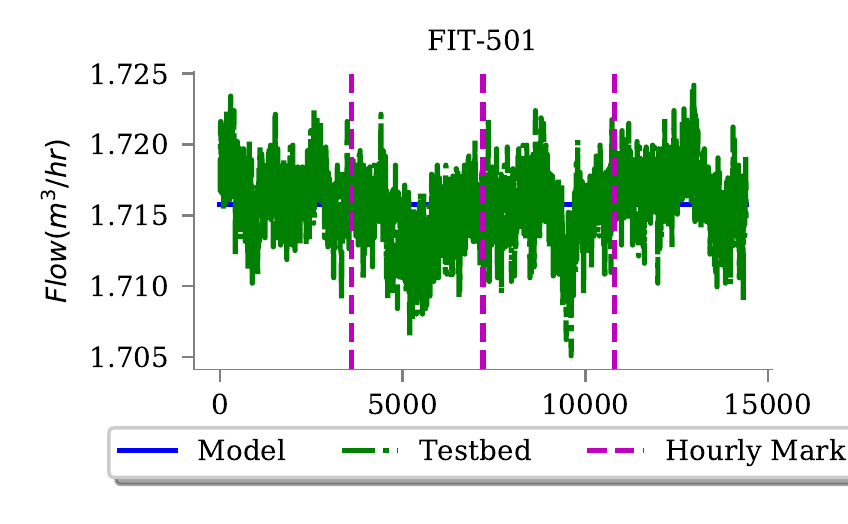}}
\subfigure[FIT-502 / RO Permeate]{\label{fig:flow_4hrs_6}\includegraphics[scale = 0.46]{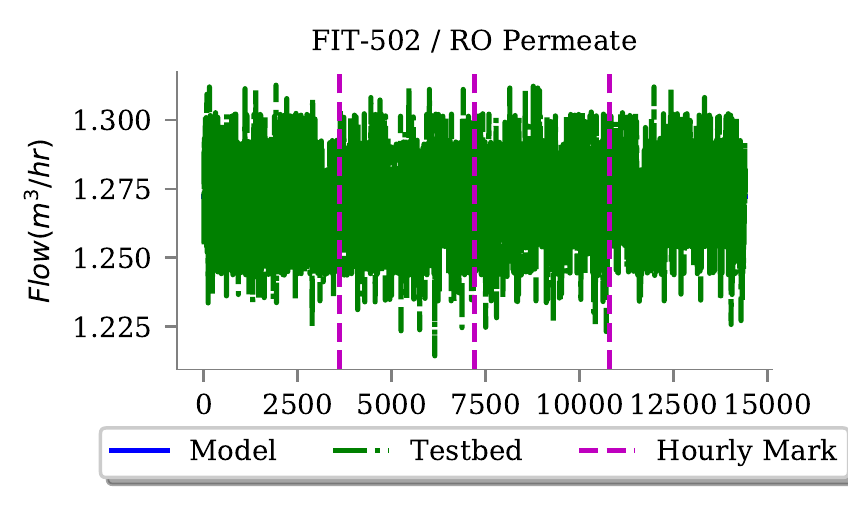}}
\caption{Comparison of Simulated and Actual Flow Over 4hrs in SWaT}
\label{fig:flow_4hrs}
\end{figure}

\begin{figure}[htbp]
\centering
\subfigure[LIT-101]{\label{fig:level_4hrs_a}\includegraphics[scale = 0.46]{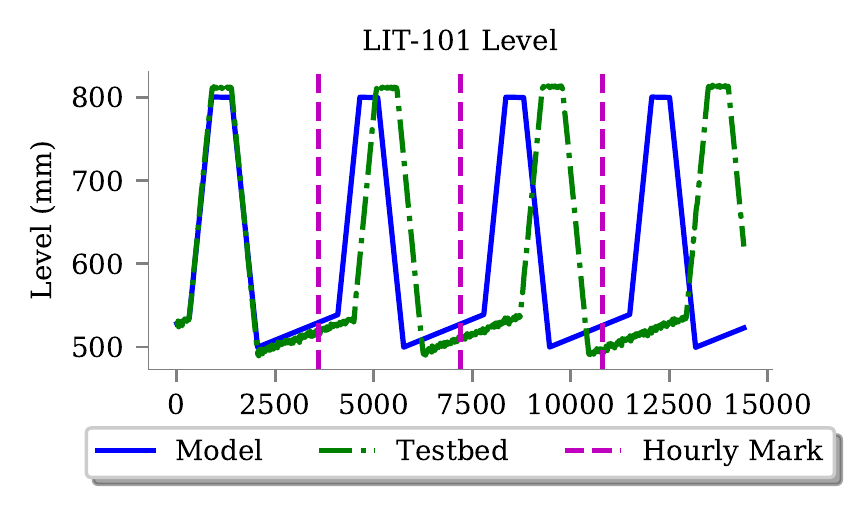}}
\subfigure[LIT-301]{\label{fig:level_4hrs_b}\includegraphics[scale = 0.46]{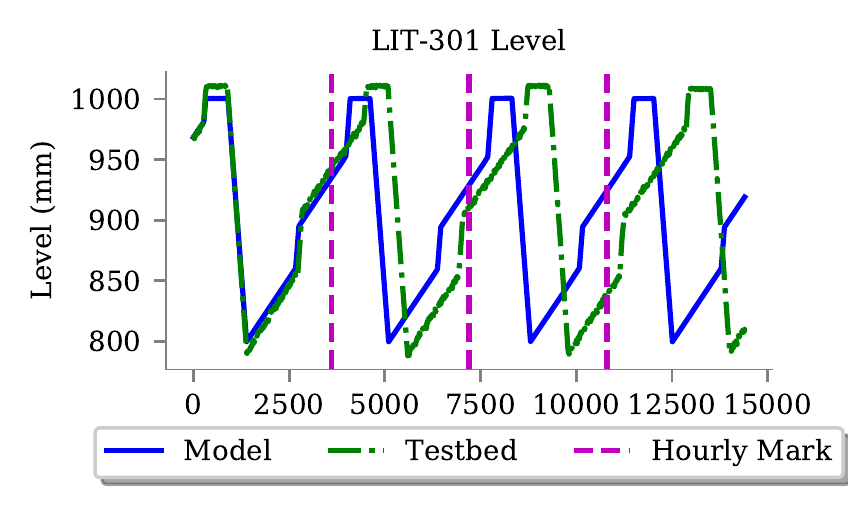}}
\subfigure[LIT-401]{\label{fig:level_4hrs_c}\includegraphics[scale = 0.46]{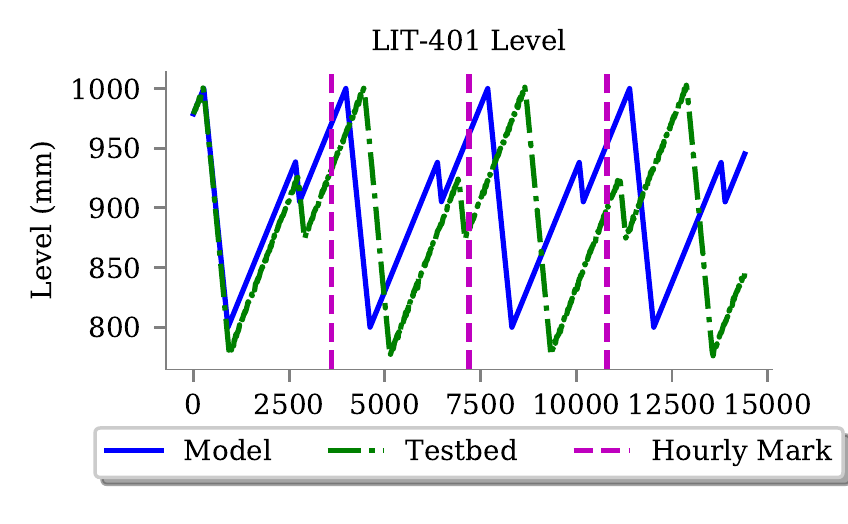}}
\caption{Comparison of Simulated and Actual Tank Levels Over 4hrs in SWaT}
\label{fig:level_4hrs}
\end{figure}

The model was first validated against historical testbed data under normal operating conditions for 4 hours. Figures \ref{fig:flow_4hrs} and \ref{fig:level_4hrs} shows that the model is able to predict the behaviour of the testbed as the trends due to control strategy of the plant were replicated. However, we observed that as simulation time increases, the historical plant data lags behind the model prediction. This was due to the ideal assumptions of the model where state-switching time (i.e. time to change valve and pump states) is instantaneous and flows were assumed to be constant at the design specifications. The time lags due to ideal assumption increases over simulation time, however, this can be mitigated by running shorter simulation cycles (about 1 hour).

\subsubsection{Attack 7 : SSSP Attack.} Attack 7 from the dataset was used to validate a single point single stage attack on the testbed. LIT-301 was spoofed to 1200mm (High High Level), triggering the controllers to close MV-201 and stopping P1 pumps. This would stop flow out of P1 and stop flow into P3, hence we would expect LIT-101 to increase and LIT-301 to decrease. 

\begin{figure}[htbp]
\centering
\subfigure[FIT-101]{\label{fig:flow_4hrs_a7_a}\includegraphics[scale = 0.46]{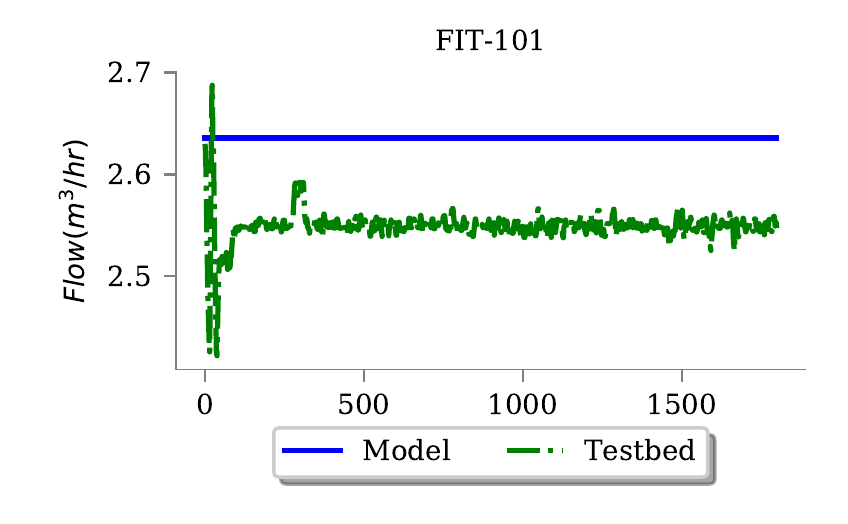}}
\subfigure[FIT-201]{\label{fig:flow_4hrs_a7_b}\includegraphics[scale = 0.46]{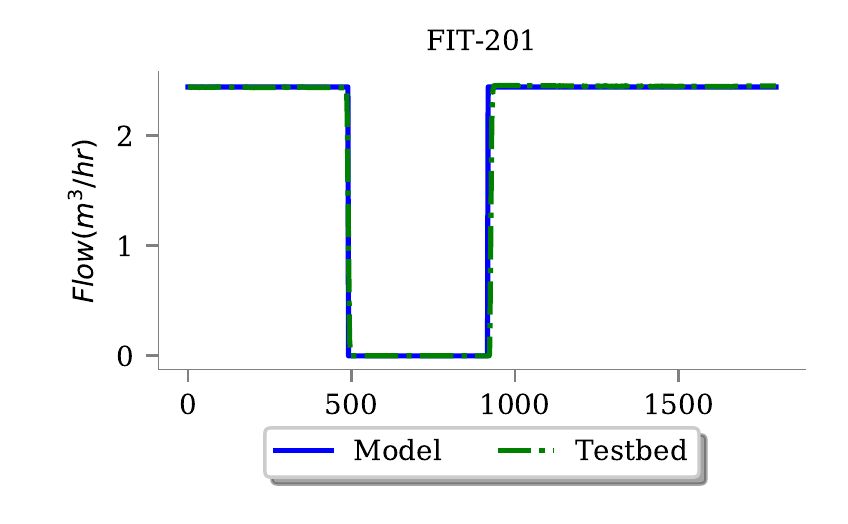}}
\subfigure[FIT-301]{\label{fig:flow_4hrs_a7_c}\includegraphics[scale = 0.46]{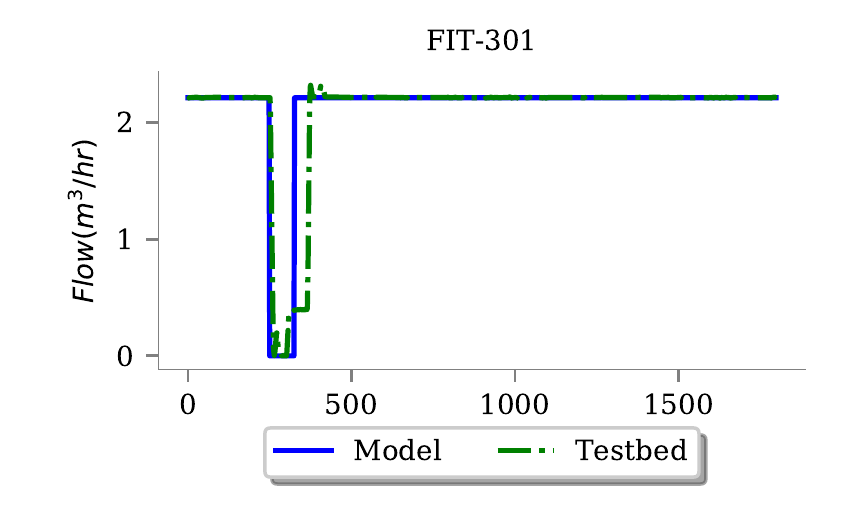}}
\subfigure[FIT-401]{\label{fig:flow_4hrs_a7_d}\includegraphics[scale = 0.46]{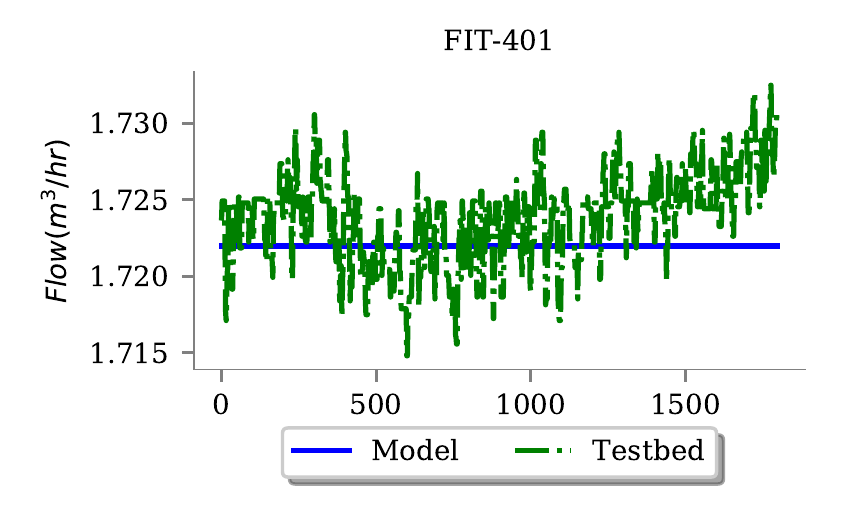}}
\subfigure[FIT-501]{\label{fig:flow_4hrs_a7_e}\includegraphics[scale = 0.46]{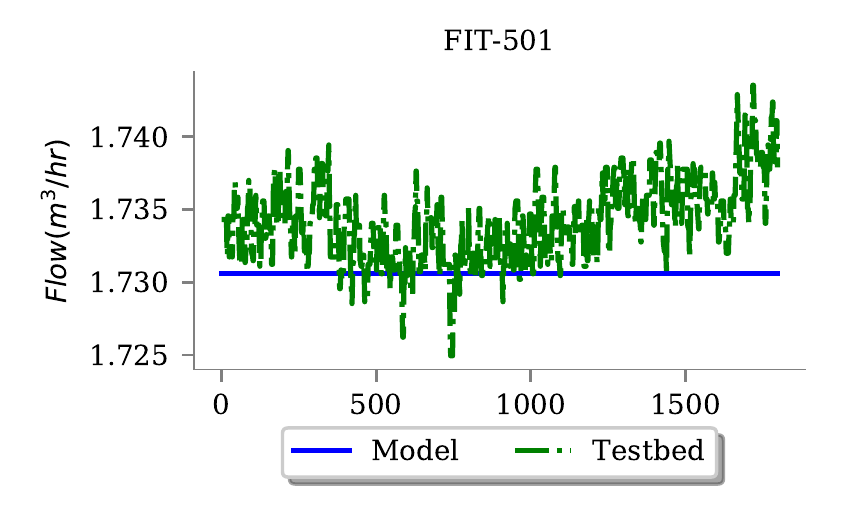}}
\subfigure[FIT-502 / RO Permeate]{\label{fig:flow_4hrs_a7_f}\includegraphics[scale = 0.46]{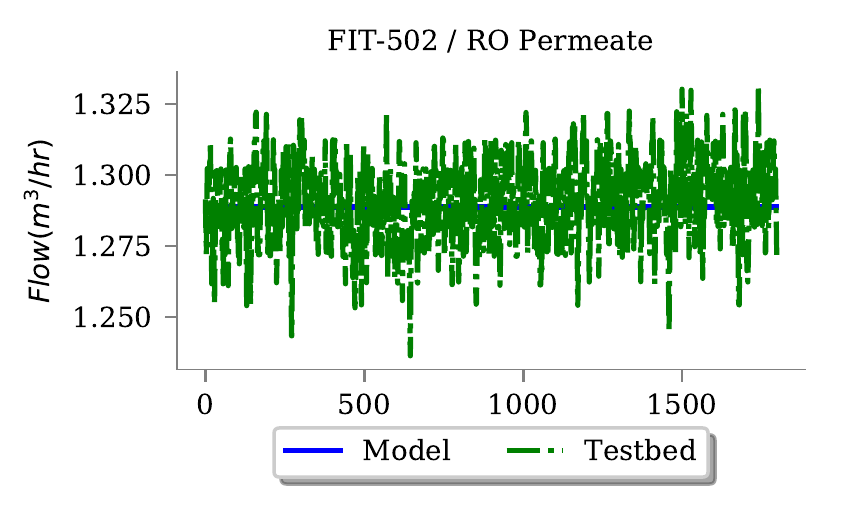}}
\caption{Summary of Flow Over Duration of SSSP Attack}
\label{fig:flow_4hrs_a7}
\end{figure}

\begin{figure}[htbp]
\centering
\subfigure[LIT-101]{\label{fig:level_4hrs_a7_a}\includegraphics[scale = 0.46]{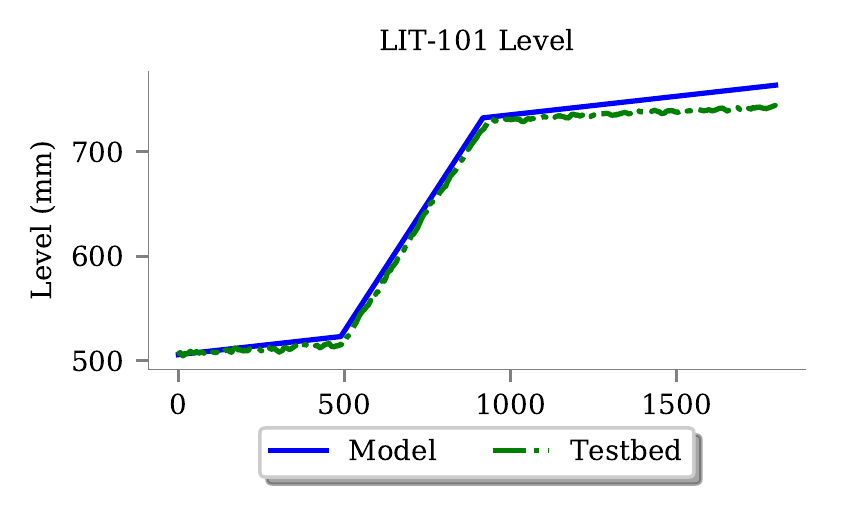}}
\subfigure[LIT-301]{\label{fig:level_4hrs_a7_b}\includegraphics[scale = 0.46]{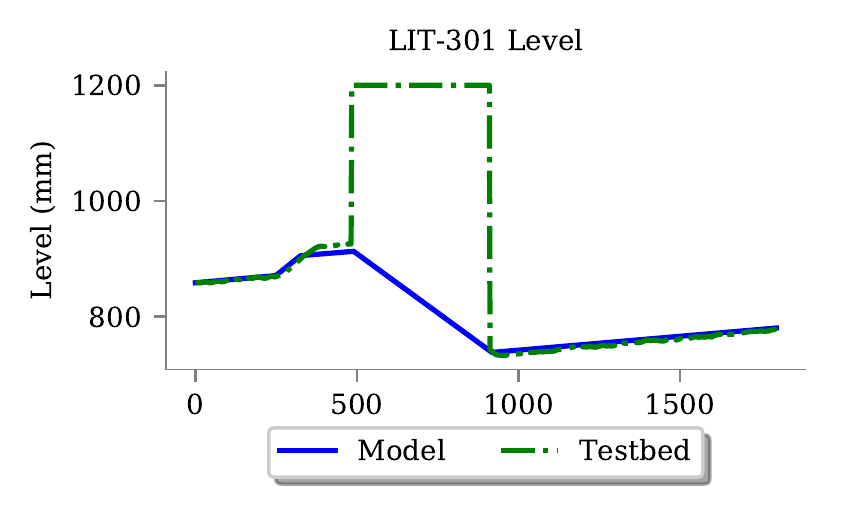}}
\subfigure[LIT-401]{\label{fig:level_4hrs_a7_c}\includegraphics[scale = 0.46]{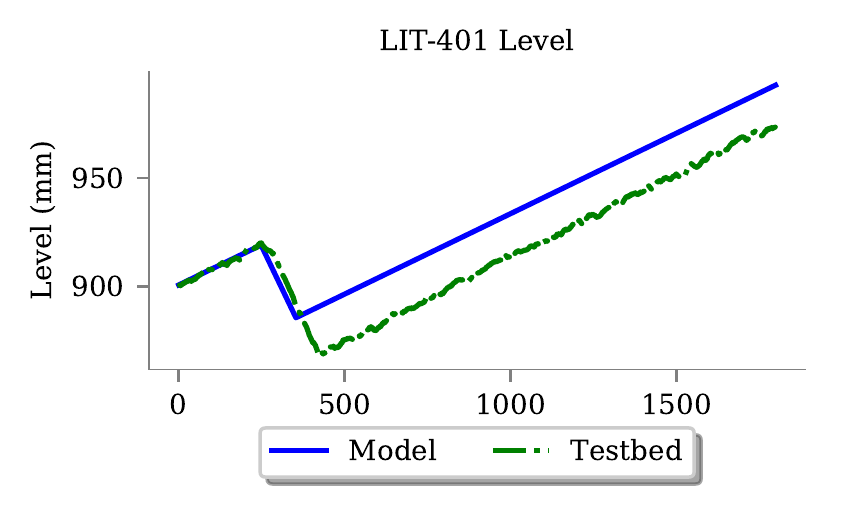}}
\caption{Summary of Tank Levels Over Duration of SSSP Attack}
\label{fig:level_4hrs_a7}
\end{figure}

From Figure \ref{fig:level_4hrs_a7_b}, it could be observed that at 500s, the attack commenced and the LIT-301 level was immediately changed to 1200mm, and during the period of 500s when the level was spoofed, the tank level decreases as no water enters P3 (no flow in FIT-201) and there is flow out of P3 (indicated by non-zero flow in FIT-301).
From Figures \ref{fig:flow_4hrs_a7_b} and \ref{fig:level_4hrs_a7_a}, FIT-201 flow rate is 0 $m^3/hr$ due to the closure of MV-201, indicating there is no flow out of P1 and hence, an increase in LIT-101 was observed. When the attack was removed, it was noted that the model prediction for tank level LIT-301 matches that of the sensor data with genuine tank level, indicating that the model was able to predict the actual physical water tank level during an attack accurately.

\subsubsection{Attack 30 : MSMP Attack.} Attack 30 from the dataset was used to validate a multi-stage,multi-point attack on the testbed. LIT-101 tank level was spoofed to a constant level of 700mm , actuator commands were spoofed to keep MV-101 closed,  P101 on and MV-201 open. The intended result was to cause Tank 1 to underflow and Tank 3 to overflow. The model is able to predict the behaviour of attack during the attack, the response of the system of Tank 1 level decreasing and Tank 3 level increasing (Appendix \ref{apx:msmp}, Figure \ref{fig:level_4hrs_a30_a} and \ref{fig:level_4hrs_a30_b} respectively) were accurately predicted. Similarly, our model was able to predict accurately  the actual physical tank level of Tank 1 (LIT-101) while the sensor level was being spoofed.

\subsection{Discussion}
In all, the proposed model construction approach based on threat intent and system design specifications was able to extract relevant control components such as sensors and actuators and build the model with dependencies based on control strategies and system design. This approach of identifying a subset of control variables and components simplifies the model for easy construction and analysis, especially for large scale systems where many components are unessential with respect to the threat model we wish to evaluate against. The model was validated and found to be able to simulate the actual plant operation with high accuracy for simulation time of less than an hour. The model allows us to simulate plant behaviour under normal and attack conditions for analysis.

\section{Sensitivity Analysis: Cascading Impact of Attacks}\label{sec:SA}
With the model validated, we proceed to conduct sensitivity analysis by performing SSSP attacks on P1 and P3 to study the cascading effect of attacks within the system from attacks that originate from a single point. For our experiments, the attacker's goal is to decrease the throughput of the system by stopping flow of water. This could be achieved by spoofing actuator commands or spoofing sensor values to trick the controllers to change states of actuators such as valves and pumps to manipulate flow of water in and out of the whole system or specific processes.

\subsection{Vulnerable and Critical States for SWaT} 
Vulnerable states are states in normal operation which are closest to critical states and are defined locally (i.e. for sub-processes), whereas critical states can be either local or global. Global critical states are states where the whole system's performance or safety has reached an unacceptable level. For our study, we investigate an attack's cascading effects with respect to global performance critical state where the throughput of the testbed has reached zero.

From the normal operating curves in Figures \ref{fig:flow_4hrs} and \ref{fig:level_4hrs}, we determine the highest and lowest operating points (i.e. 500mm and 800mm for Tank 1 and 800mm and 1000mm for all other tanks) as the vulnerable states as they are states closest to the local critical states such as overflow or underflow of tanks. We use the system state at these specific vulnerable states for the model initial conditions for sensitivity analysis of data spoofing attacks.

\subsection{Time-to-Critical-State from Attack Commencement} The time-to-critical-state is the measure of distance of a system to an unacceptable state. For our experiments, we measure the time-to-critical-state of the system from the commencement of the attack where the attack location is a local vulnerable state. The time-to-critical-state would be the time from attack when there is no flow measured or when there is a safety violation such as overflow or underflow of tank. 

The worst-case time-to-critical-state of the system is the shortest time required to bring a system to a local unacceptable state, where in our case, it would be either to overflow or underflow a tank. We subsequently computed the worst-case time-to-critical-states (Table \ref{table:ttcs}) from the identified vulnerable states of tank levels. The time-to-critical-states for flow (i.e. zero flow) would be dependent on states of actuators that control the flow, which could be manipulated by other control variables in the control loop.

\begin{table}[htbp]
\centering
\begin{tabular}{|l|c|c|c|}
\hline
\multirow{2}{*}{}              & \multicolumn{3}{l|}{\textbf{Time-To-Critical-State (s)}} \\ \cline{2-4} 
                               & \textbf{P1}       & \textbf{P3}       & \textbf{P4}      \\ \hline
\textbf{High Vulnerable State} & 838.8             & 448.1             & 493.3           \\ \hline
\textbf{Low Vulnerable State}  & 1101.4            & 1939.8            & 2523.2           \\ \hline
\end{tabular}
\caption{Summary of worst-case time-to-critical-states of tank levels}
\label{table:ttcs}
\end{table}

\subsection{Approach for Analysis of Cascading Effects} The cascading effects of attacks would be measured for each process by: 1) computing the impact ratio of the tank levels and flow rates, and 2) determining the time-to-critical state from the commencement of an attack. The local impact of attack would allow us to understand the propagation of attack from a specific location to other parts of the system.

\subsection{Attack Experiments} The goal of the experiments is to study the effects of integrity attacks on the system state in a single stage and its effects on the other 4 stages. The attacks can be categorized to 1) stopping water inflow into the system and 2) disrupting flow in the system. By stopping the inflow of water into system, we investigate how the disruption of inflow of water can be achieved by integrity attacks and the propagation of attack impact based on how the sub-systems are physically connected. In the second experiment to disrupt the flow of water by manipulating pump states through level sensor values, we investigate the exploitation of control inter-dependencies and how attack impact propagates via control dependencies. In all, we studied the cascading impact of attacks due to how the sub-systems are connected either physically or through control dependencies.  

\subsection{Stopping Water Inflow Into System} We first explore the effects of attackers conducting attacks to stop inflow of water into the system. This can be achieved by: a) spoofing actuator commands to close inlet valve (MV-101) and, b) spoofing P1 tank level to "High" to trick the controller to close the valve. The attacks were conducted at the system's first High Vulnerable State and Low Vulnerable State where Tank 1 is at the highest and lowest normal operating level respectively. The location of attack is MV-101, the valve that controls water inflow into the system and the flow is measured by FIT-101.

\begin{figure}[h]
\centering
\subfigure[Impact Ratio]{\label{fig:flow_exp1_IR}\includegraphics[scale = 0.8]{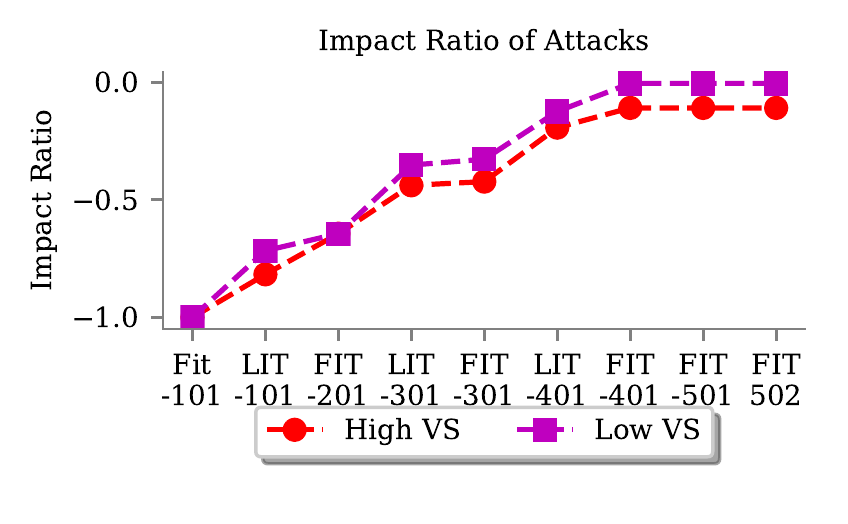}}
\subfigure[Time-to Critical-State]{\label{fig:flow_exp1_TTCS}\includegraphics[scale = 0.8]{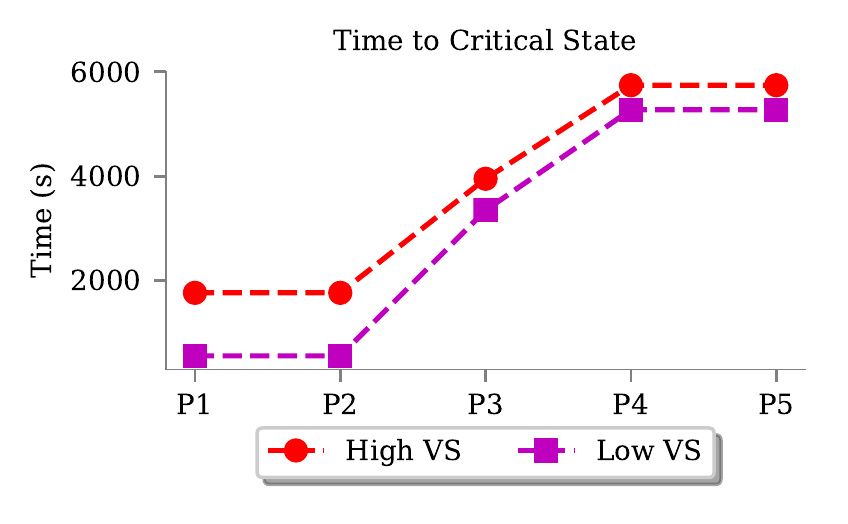}}
\caption{Metrics for Impact of Attacks That Closes MV-101}
\label{fig:exp1_metric}
\end{figure}

It was observed that the attack by command injection to turn off MV-101 and spoofing of tank level to "High" state did not result in any difference in the impact on the system. The reason was that the control loop that controls MV-101 acts independently and do not have any other control variables other than LIT-101. As such, the spoofing of tank level was equivalent to injecting command to turn off the valve. Operating curves of system under normal and attack scenarios can be found in Appendix \ref{apx: expt} Figures  \ref{fig:flow_exp1} and \ref{fig:level_exp1}.

\subsubsection{Impact Ratio.} The Impact Ratio (Figure \ref{fig:flow_exp1_IR}) of the operational metrics at the end of the attack shows the magnitude of the attack on the operation. It was found that for this particular attack, the highest impact was at the location of the attack and all other ratios were negative, indicating that the decrease in operational performance is greatest at the point of attack and the magnitude of impact diminishes downstream. The impact of attack was found to be greater when the initial state of the system was at the high vulnerable state as compared to the low vulnerable state, this is due to the additional capacity in the tank which has to be emptied to result in no flow.

\subsubsection{Time-to-critical-state.} The time-to-critical-state is the time taken for each process to reach an unacceptable state. From Figure \ref{fig:flow_exp1_TTCS}, it was observed the time to reach critical state increases downstream, which was expected from processes that are connected sequentially. For the attack when initial state was at the low vulnerable state, the time-to-critical state was found to be shorter, implying that the system reaches unacceptable state in a shorter time and consequence of attack is achieved quicker. It was also observed that the time-to-critical-state were identical for pairs of processes which do not have tanks (i.e. P1-P2 and P4-P5), whereas there is an increase in the time for processes with tanks. This implies that the tanks act as buffer against the impact of attacks.

\subsection{Disrupting Flow in System by Manipulating Pump State} The effects of manipulating pump states on the system were explored. Pumps could be switched off by: a) command spoofing and, b) spoofing tank levels to trick the controller to send commands to change the pump state. We conduct these attacks separately for P1 and P3, analysis of the system's resilience to such attacks were performed.

\subsubsection{Attack on P1 pumps.} The location of the attack is at the pumps of P1 where water is removed from Tank 1 (measured by LIT-101) and flow into P2 is measured by FIT-201. The following 3 attacks were conducted: 1) Command spoofing to switch off pumps, 2) Spoof Tank 1 state to "Low Low", and 3) Spoof Tank 3 state to "High". Operating curves of system under normal and attack scenarios can be found in Appendix \ref{apx: expt} Figures \ref{fig:flow_exp2_p1} and \ref{fig:level_exp2_p1}.

\begin{figure}[htbp]
\centering
\subfigure[Impact Ratio]{\label{fig:flow_exp2_p1_ir}\includegraphics[scale = 0.8]{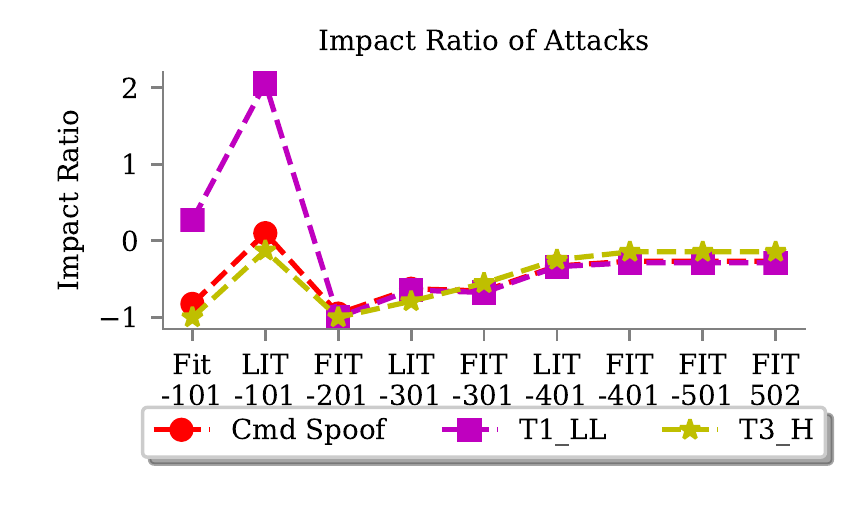}}
\subfigure[Time-to Critical-State]{\label{fig:flow_exp2_p1_ttcs}\includegraphics[scale = 0.8]{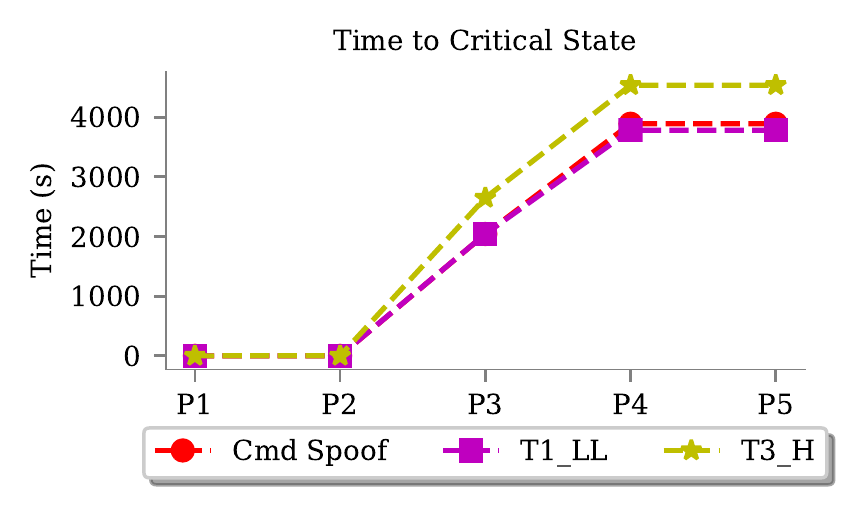}}
\caption{Metrics for Impact of Attacks That Turns Off Pumps in P1}
\label{fig:metric_exp2_p1}
\end{figure}

It could be observed from Figure \ref{fig:flow_exp2_p1_ttcs} that the effect of attacks on P1 and P2 were instantaneous and flow was disrupted. However, due to the tank in P3, disruption of flow was only observed after the tank was depleted. From Figure \ref{fig:flow_exp2_p1_ir}, the impact of attack for command spoofing and spoofing of Tank 3 level were similar where the impact was largest for FIT-101 and FIT-201. The spoofing of Tank 1 level resulted in a positive ratio due to the increase in tank level from the exploitation of control actions to maintain inflow into tank and stop outflow. 

\subsubsection{Attack on P3 pumps.} The location of the attack is at the pumps of P3 where water is removed from Tank 3 (measured by LIT-301) and flow into P4 is measured by FIT-301. The following 3 attacks were conducted: 1) Command spoofing to switch off pumps, 2) Spoof Tank 3 state to "Low Low", and 3) Spoof Tank 4 state to "High". Operating curves of system under normal and attack scenarios can be found in Appendix \ref{apx: expt} Figures \ref{fig:flow_exp2_p3} and \ref{fig:level_exp2_p3}.

\begin{figure}[htbp]
\centering
\subfigure[Impact Ratio]{\label{fig:flow_exp2_p3_ir}\includegraphics[scale = 0.8]{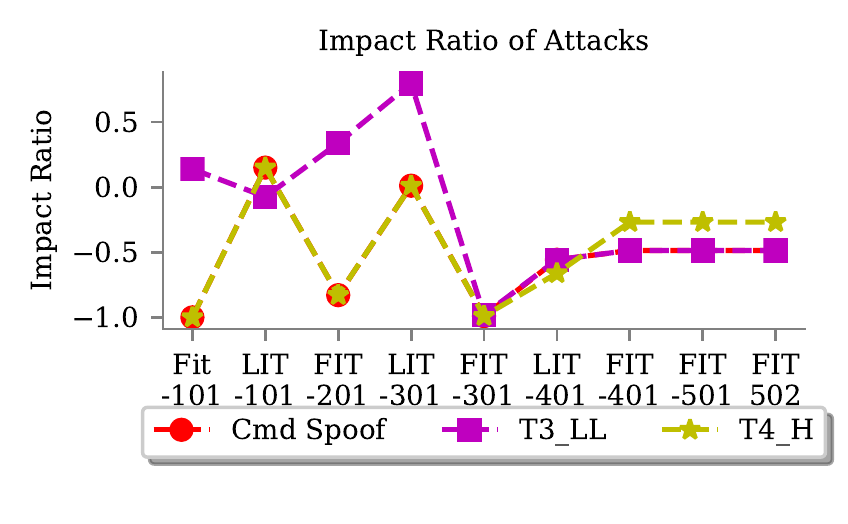}}
\subfigure[Time-to Critical-State]{\label{fig:flow_exp2_p3_ttcs}\includegraphics[scale = 0.8]{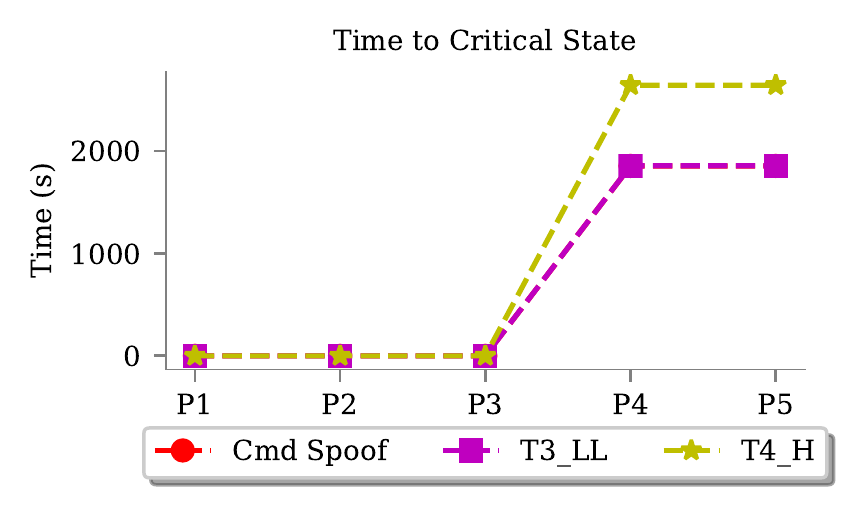}}
\caption{Metrics for Impact of Attacks That Turns Off Pumps in P3}
\label{fig:metric_exp2_p3}
\end{figure}

From Figure \ref{fig:flow_exp2_p3_ttcs}, it could be observed that the effect of attacks on P3 and its preceding processes where  flow was instantaneously halted in these processes (time-to-critical-state is 0). The time taken for P4 to reach critical state was attributed to the presence of Tank 4, where its contents had to be depleted. When flow in P4 is disrupted, the downstream process P5 was immediately affected. From the impact ratio in Figure \ref{fig:flow_exp2_p3_ir}, it was found that the impact of attack for command spoofing and spoofing of Tank 4 level were identical where water flow from P1 to P3 were disrupted. The spoofing of Tank 3 level exploited the control actions that resulted in continuous inflow into Tank 3 and no outflow, which could be observed from the positive ratio.

\subsection{Recommendation of Design Improvements}
Various attacks were simulated on the model of the testbed and the behaviour of the system under attack was analyzed. The cascading effects of attacks on each process were quantified using the Impact Ratio and the time taken for each process to reach critical state was measured. The Impact Ratio uses the difference in area-under-curve of operating curves, where the operating function is the performance variable of any CPS component such as tank water level or flow throughput in a water treatment plant and for a power plant, the power generated by a generator. The time-to-critical-state is the time taken for a system to reach a defined critical state and it is used as a metric to compare a system's resilience to various attack based on their time-to-critical-state response.

It was found that processes with tanks have increased ability to withstand the attack as the time-taken to reach critical state increases in these processes. The tanks are stored capacity and act as buffer, where normal operation can continue until the tanks are emptied. To improve the overall system's ability to withstand attacks, it is recommended that tanks should be installed for every process to maximize the buffer to increase the overall time taken to reach a critical state, providing operators additional time to respond to attacks. While the use of buffers in critical system is intuitive and, it is an important and sometimes an overlooked solution. Buffers such as use of batteries in power grids or critical infrastructure would ensure continued critical operation while remediation of a fault or attack takes place. The use of buffers should be considered in totality of the improved resilience and costs of implementation.

By comparing Figure \ref{fig:flow_exp2_p1_ttcs} and \ref{fig:flow_exp2_p3_ttcs}, it was observed that the attack on P3, a later process resulted in shorter time-to-critical-state. This suggests that for sequentially connected processes, any attacks on later processes would result in decreased capability to withstand the disruptive effects of attacks, decreasing the time available for operators to respond. With this knowledge, redundancy could be introduced for later processes which are critical to operational performance in order to improve the resiliency of the system. However, for systems that are connected in a more complex manner, there is a need to further study the resilience characteristics of the system using the proposed metrics.

\vspace{-0.2cm}
\section{Conclusion and Future Work} \label{sec:conclusion}
Understanding the interaction of the adversary with the physical system and studying the behaviour of system's response to cyber attacks can help in the design of defences that improves the system's resilience. To this end, a novel approach that uses a threat model with an adversary intent and capabilities was used to determine relevant control variables in a CPS. These control variables, together with their related control strategies and design specifications were used to build a model of the CPS. The resulting model was used to study cascading effects of attacks and analyze the resilience of the CPS to data-oriented attacks.

The proposed modeling approach was demonstrated on an actual water treatment testbed and the model was validated with the testbed historical data. The model was determined to be accurate in simulating the testbed's behaviour under normal and attack scenarios. Attacks with the goal to disrupt water flow through the testbed were simulated and the impact of attack was quantified by measuring the Impact Ratio and time taken to reach critical state. The cascading impact of attacks within the system for the various attacks were analyzed and design enhancements were proposed to address identified weaknesses to improve the system's resilience to cyber attacks. 

Although the modeling approach may work well for a simple CPS such as SWaT, there might be challenges with modeling more complex and large scale CPS due to the increased inter-dependencies of subsystems and non-linear dynamics. This could potentially be resolved by using composition of smaller models of subsystems and using linear approximation for the non-linearity in dynamics. It remains as part of future work for the evaluation of the applicability of modeling approach on other CPSs (i.e. Electric Power Grid Systems where generators, switches/breakers and batteries are analogous to pumps, valves and tanks).

Further work also remains for the water treatment testbed model to study the improvement in resilience from implementing the recommended design changes. The current work looks into the cascading effects of attacks within a system, however, CPSs are usually not independent systems but are connected systems-of-systems (i.e. water treatment plant is connected to a water distribution network). Future research could focus on evaluating cascading effects of attacks on inter-connected systems.

\section{Acknowledgements}
This work was supported by the SUTD start-up research grant SRG-ISTD-2017-124. The first author's work was done during his visit to SUTD.

%
\bibliographystyle{ACM-Reference-Format}
\bibliography{biblio}

\balance

\appendix
\section{Operating Curves for Validation of Attack 30 (MSMP)} \label{apx:msmp}

\begin{figure}[H]
\centering
\subfigure[FIT-101]{\label{fig:flow_4hrs_a30_a}\includegraphics[scale = 0.46]{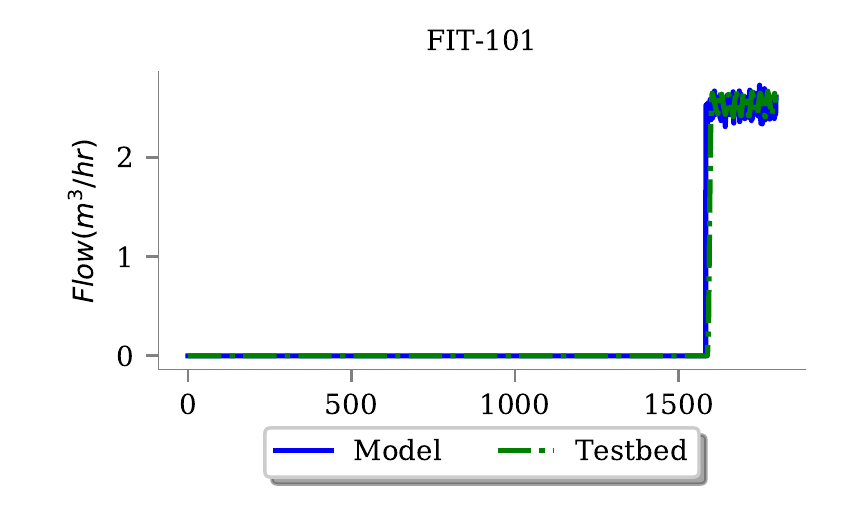}}
\subfigure[FIT-201]{\label{fig:flow_4hrs_a30_b}\includegraphics[scale = 0.46]{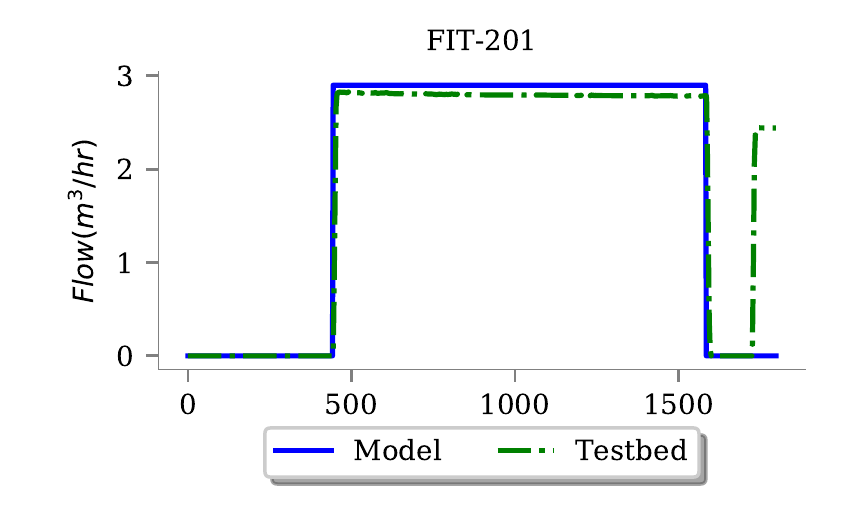}}
\subfigure[FIT-301]{\label{fig:flow_4hrs_a30_c}\includegraphics[scale = 0.46]{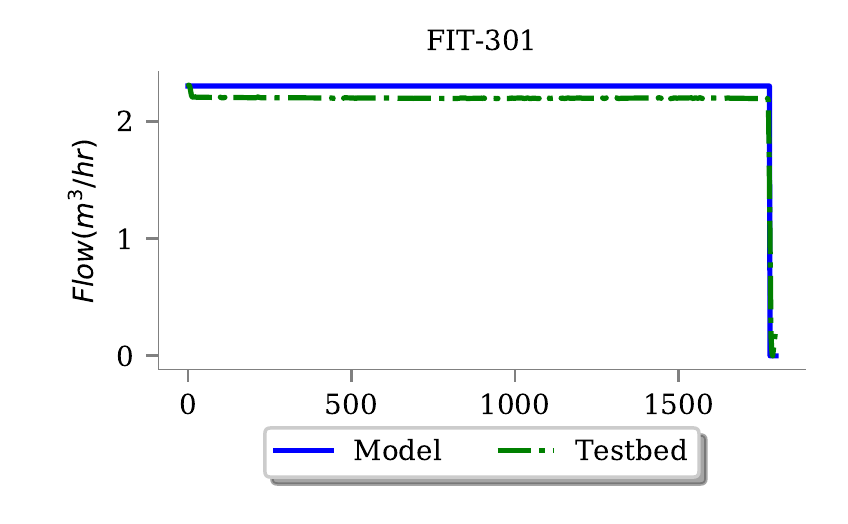}}
\subfigure[FIT-401]{\label{fig:flow_4hrs_a30_d}\includegraphics[scale = 0.46]{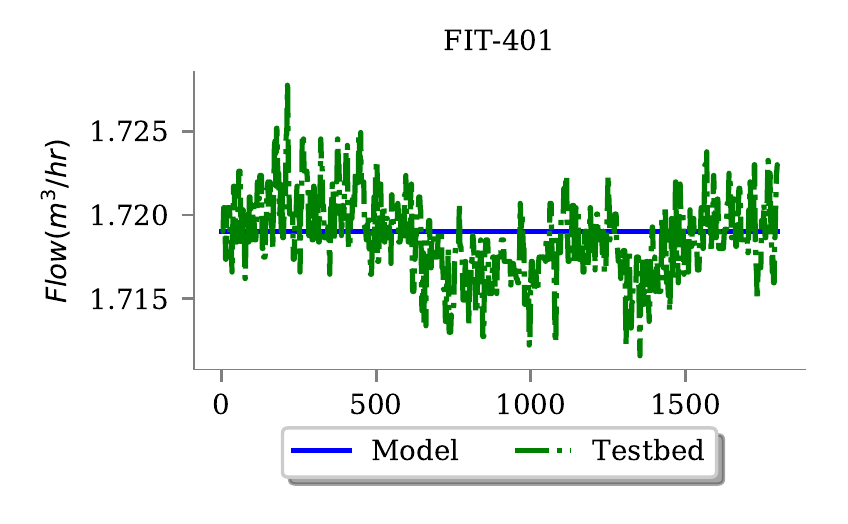}}
\subfigure[FIT-501]{\label{fig:flow_4hrs_a30_e}\includegraphics[scale = 0.46]{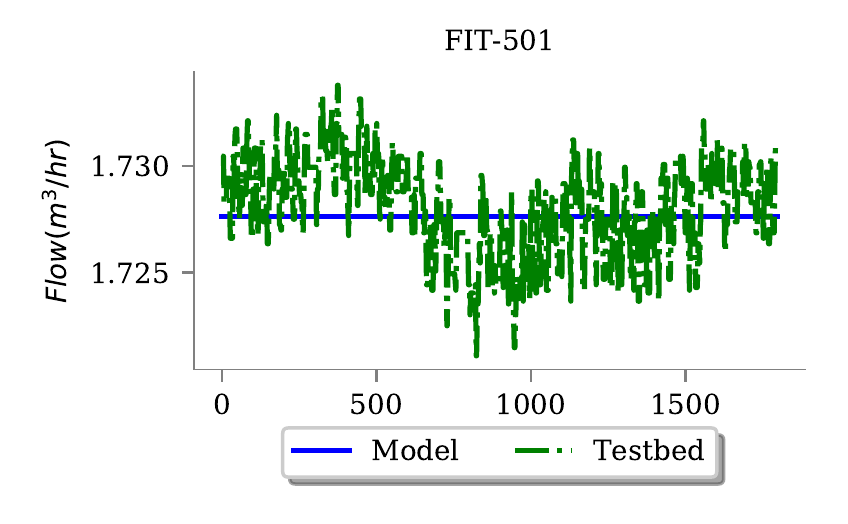}}
\subfigure[FIT-502 / RO Permeate]{\label{fig:flow_4hrs_a30_f}\includegraphics[scale = 0.46]{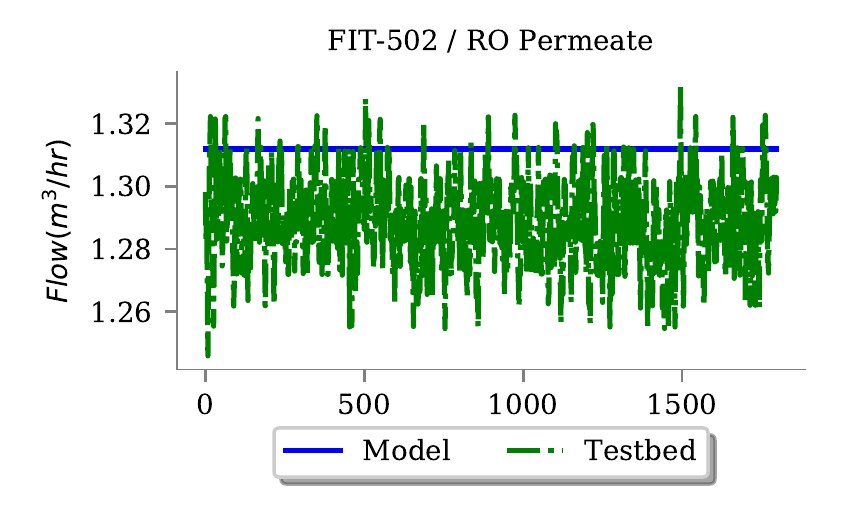}}
\caption{Summary of Flow Over Duration of MSMP Attack}
\label{fig:flow_4hrs_a30}
\end{figure}

\begin{figure}[H]
\centering
\subfigure[LIT-101]{\label{fig:level_4hrs_a30_a}\includegraphics[scale = 0.46]{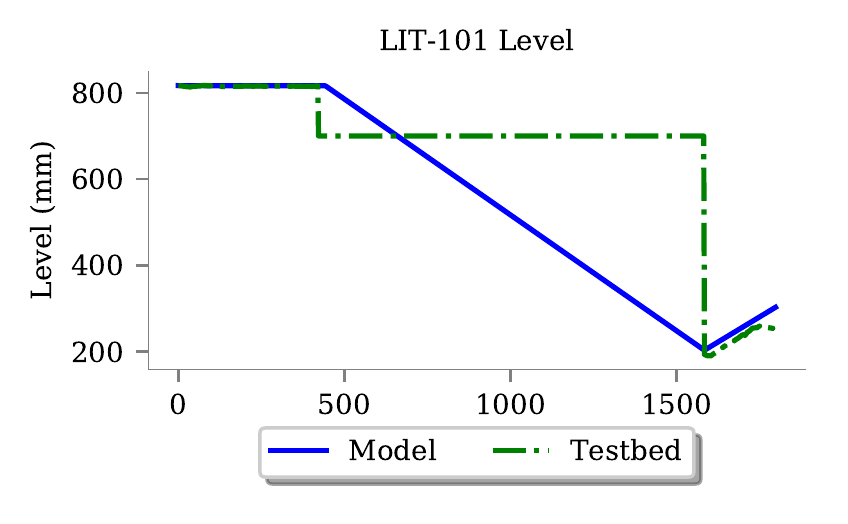}}
\subfigure[LIT-301]{\label{fig:level_4hrs_a30_b}\includegraphics[scale = 0.46]{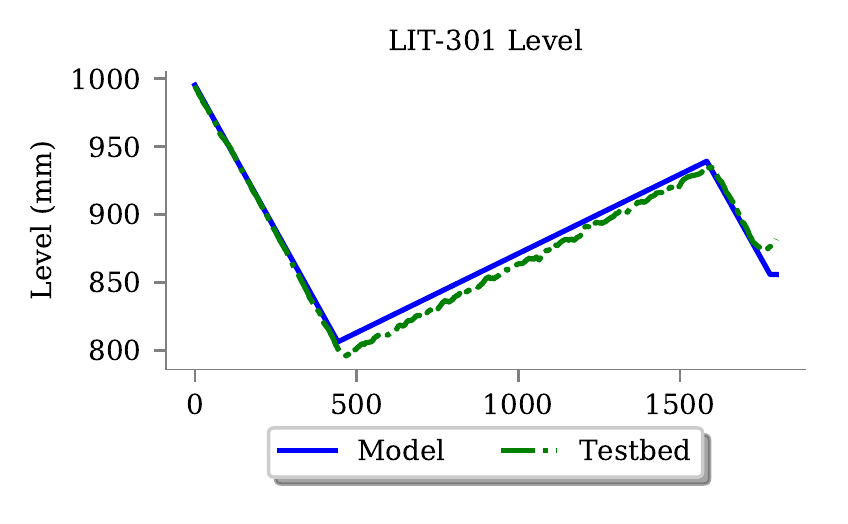}}
\subfigure[LIT-401]{\label{fig:level_4hrs_a30_c}\includegraphics[scale = 0.46]{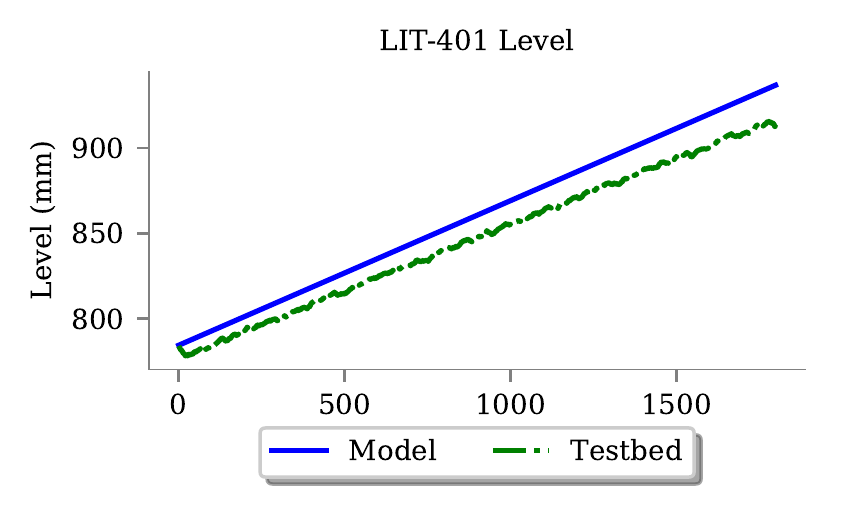}}
\caption{Summary of Tank Levels Over Duration of MSMP Attack}
\label{fig:level_4hrs_a30}
\end{figure}

Figure \ref{fig:flow_4hrs_a30} and \ref{fig:level_4hrs_a30} shows the simulation results from the model and the actual readings of flow and level sensors from the SWaT testbed during the attack period.

\section{Operating Curves for Attacks to Disrupt Water Flow} \label{apx: expt}

\begin{figure}[H]
\centering
\subfigure[FIT-101]{\label{fig:flow_exp1_a}\includegraphics[scale = 0.46]{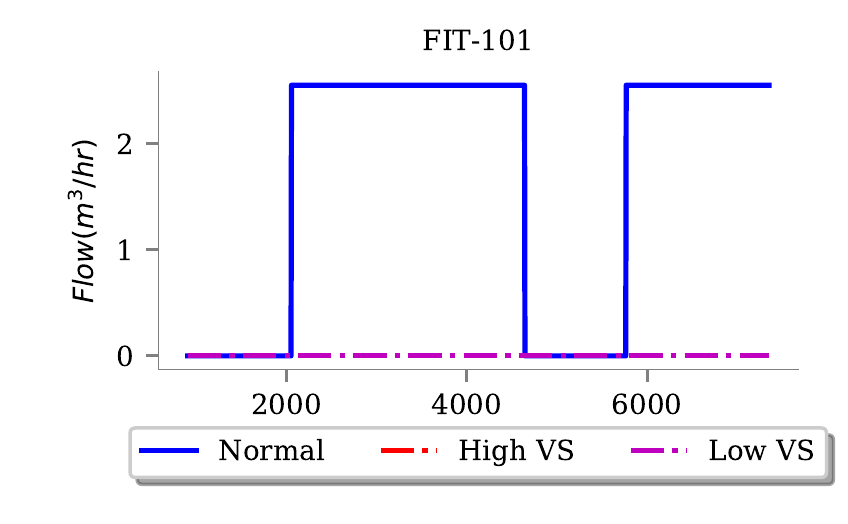}}
\subfigure[FIT-201]{\label{fig:flow_exp1_b}\includegraphics[scale = 0.46]{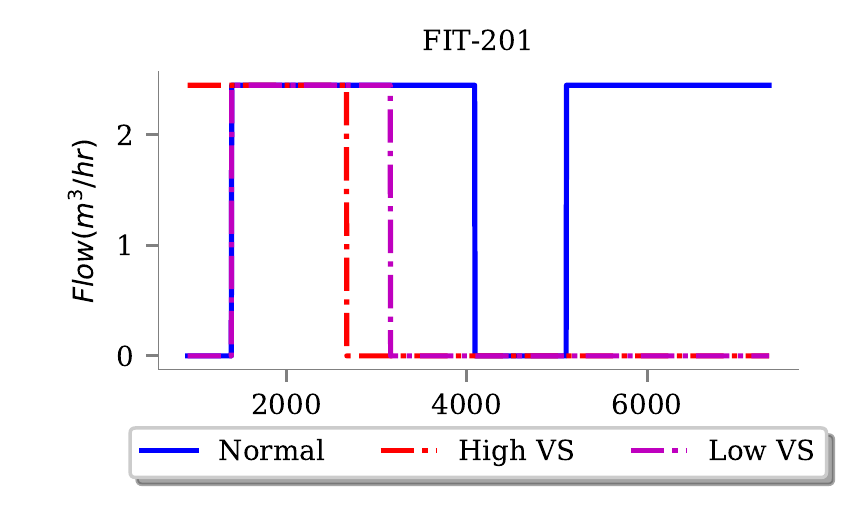}}
\subfigure[FIT-301]{\label{fig:flow_exp1_c}\includegraphics[scale = 0.46]{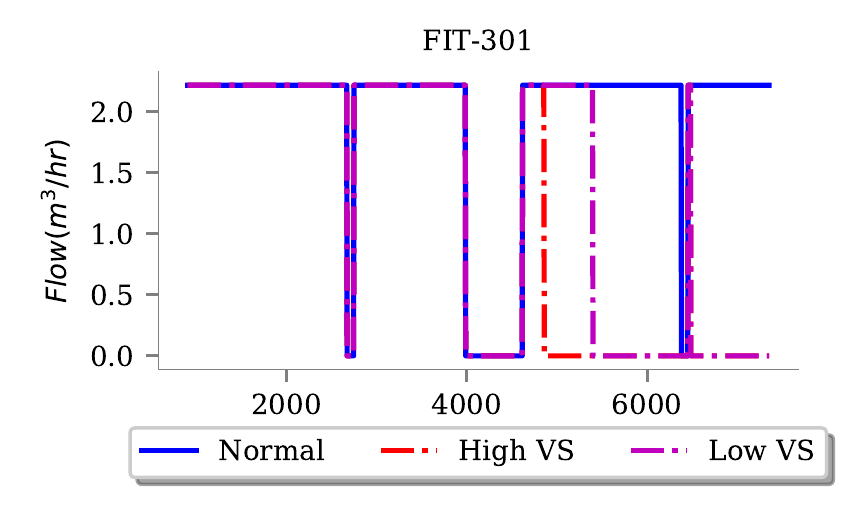}}
\subfigure[FIT-401]{\label{fig:flow_exp1_d}\includegraphics[scale = 0.46]{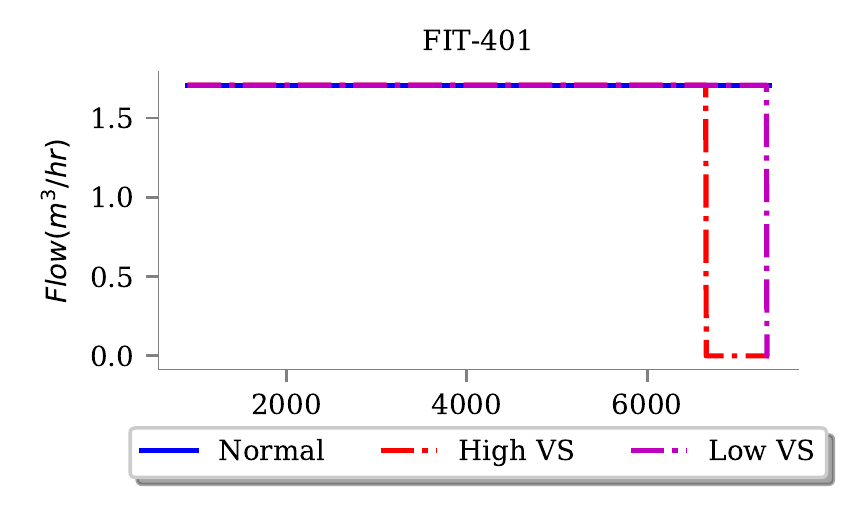}}
\subfigure[FIT-501]{\label{fig:flow_exp1_e}\includegraphics[scale = 0.46]{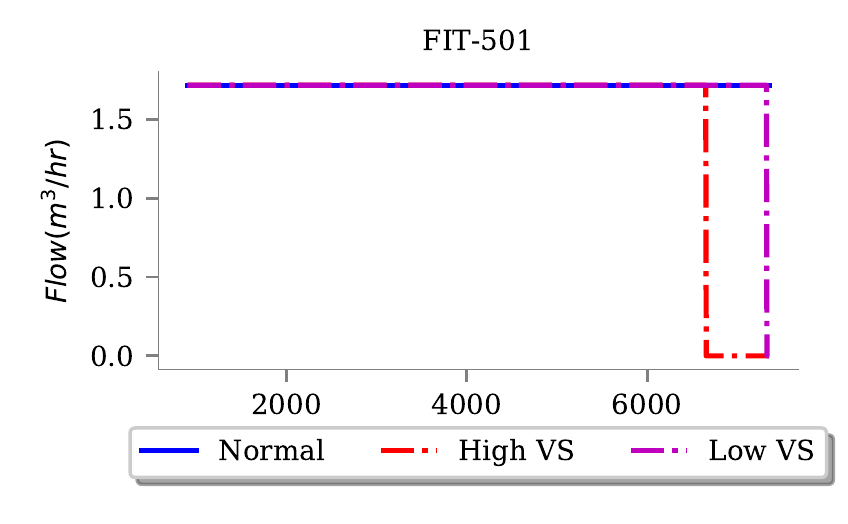}}
\subfigure[FIT-502 / RO Permeate]{\label{fig:flow_exp1_f}\includegraphics[scale = 0.46]{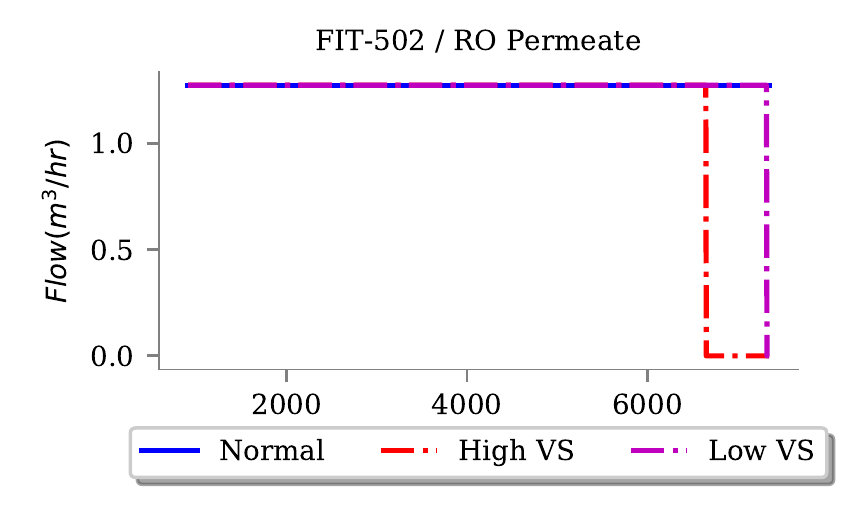}}
\caption{Summary of Flow Over Duration of Attack to Close MV-101}
\label{fig:flow_exp1}
\end{figure}

\begin{figure}[H]
\centering
\subfigure[LIT-101]{\label{fig:level_exp1_a}\includegraphics[scale = 0.46]{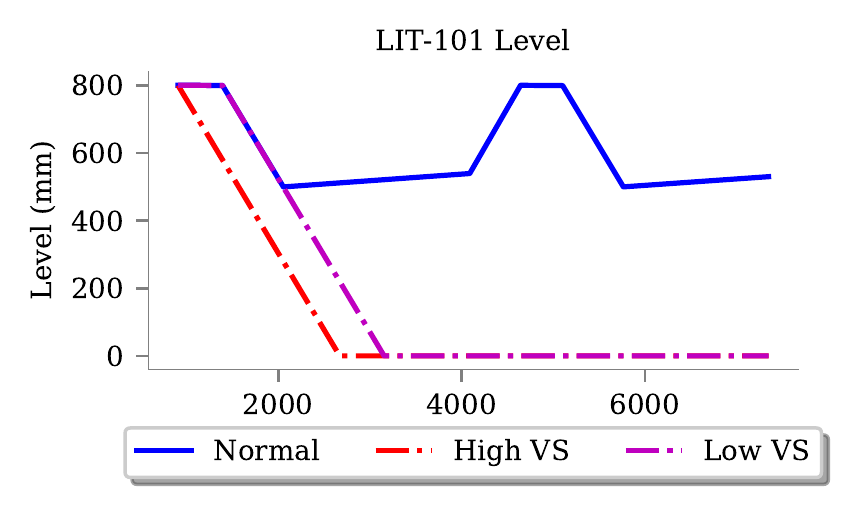}}
\subfigure[LIT-301]{\label{fig:level_exp1_b}\includegraphics[scale = 0.46]{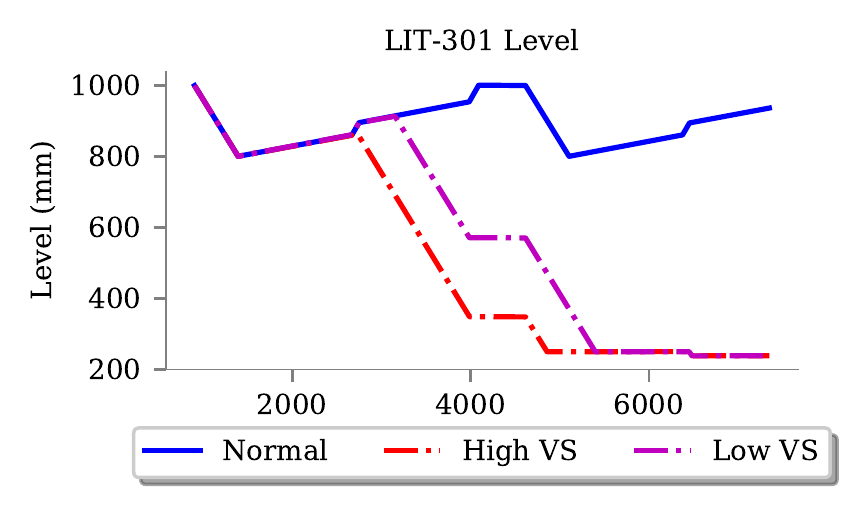}}
\subfigure[LIT-401]{\label{fig:level_exp1_c}\includegraphics[scale = 0.46]{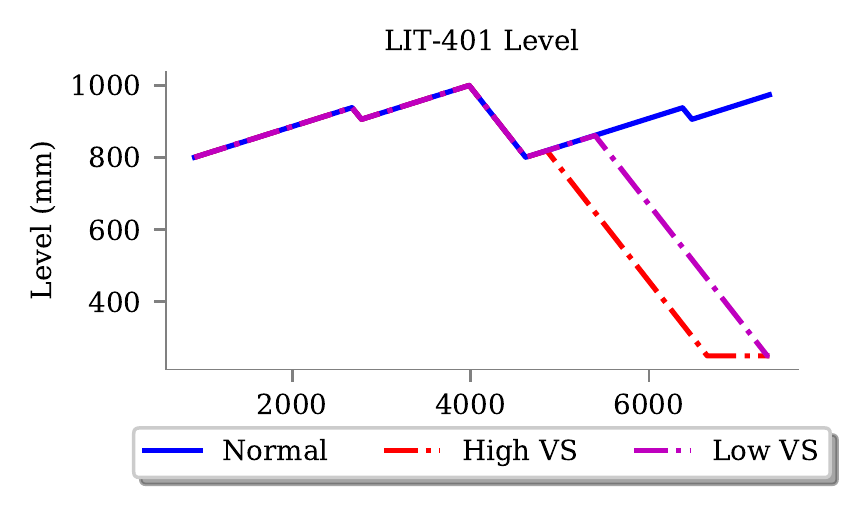}}
\caption{Summary of Tank Levels Over Duration of Attack to Close MV-101}
\label{fig:level_exp1}
\end{figure}

Figures \ref{fig:flow_exp1} and \ref{fig:level_exp1} shows the simulation result for the flow and level sensor measurements for the experiment to close MV-101.

\begin{figure}[h]
\centering
\subfigure[FIT-101]{\label{fig:flow_exp2_p1_a}\includegraphics[scale = 0.46]{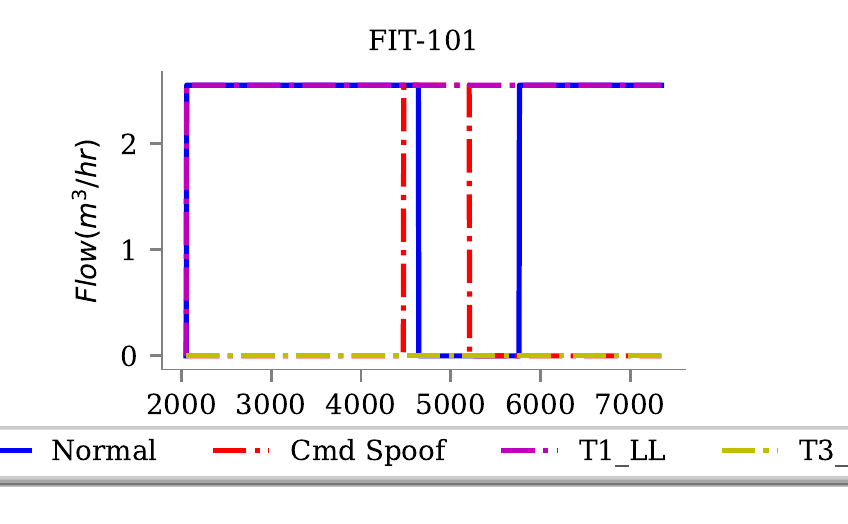}}
\subfigure[FIT-201]{\label{fig:flow_exp2_p1_b}\includegraphics[scale = 0.46]{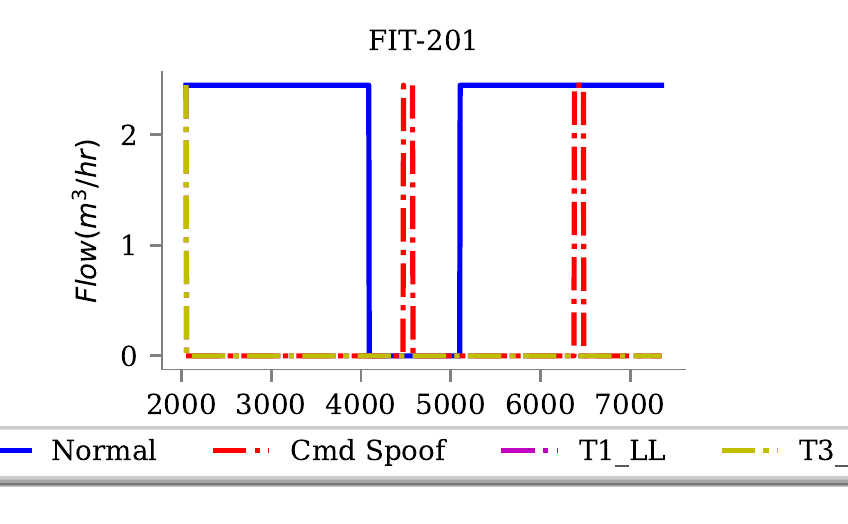}}
\subfigure[FIT-301]{\label{fig:flow_exp2_p1_c}\includegraphics[scale = 0.46]{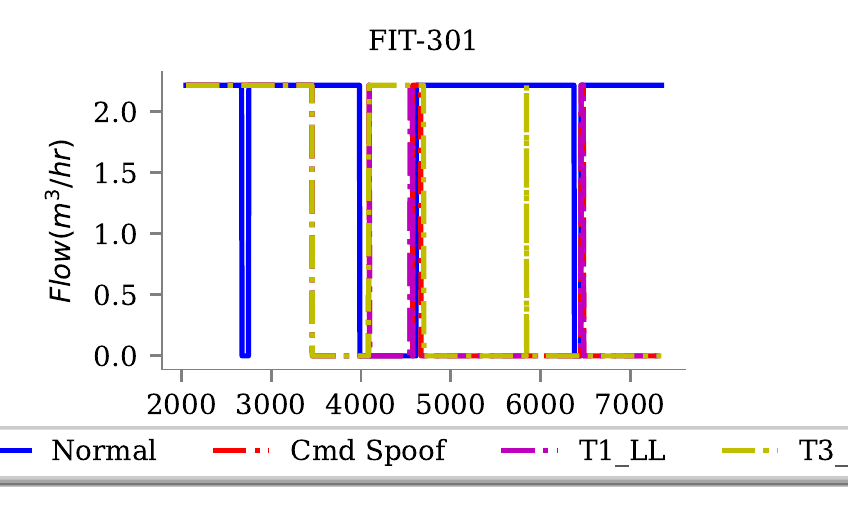}}
\subfigure[FIT-401]{\label{fig:flow_exp2_p1_d}\includegraphics[scale = 0.46]{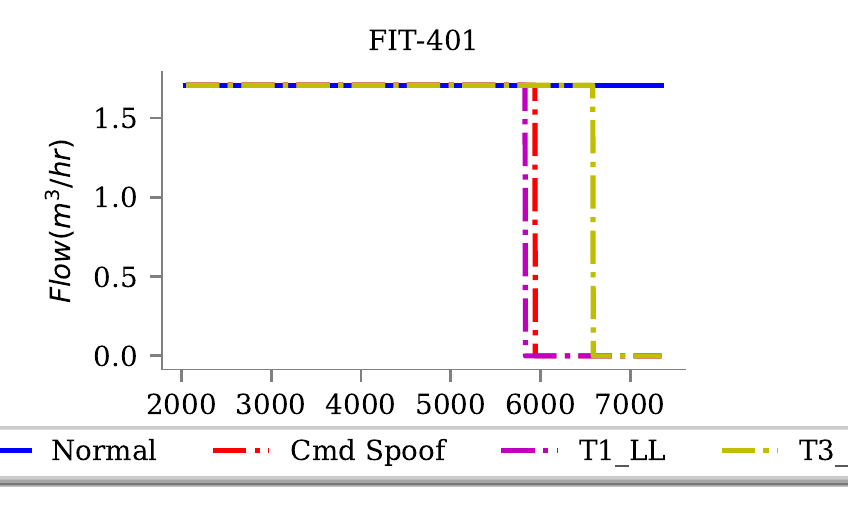}}
\subfigure[FIT-501]{\label{fig:flow_exp2_p1_e}\includegraphics[scale = 0.46]{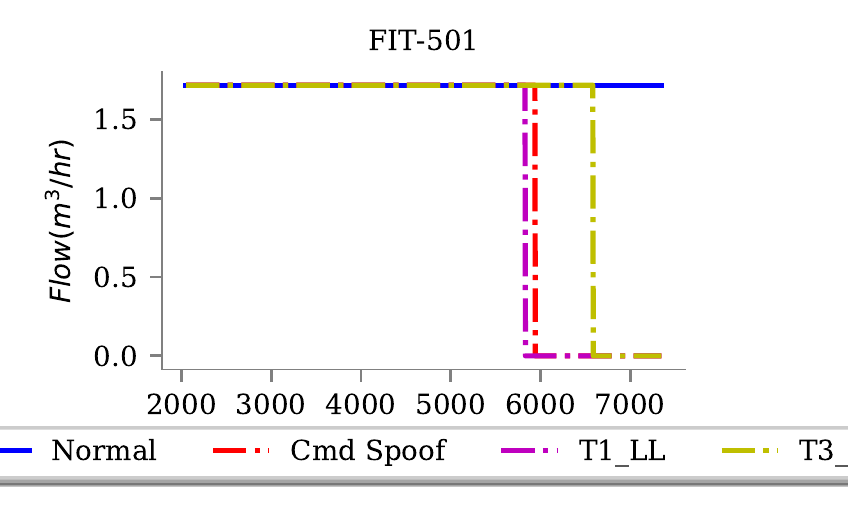}}
\subfigure[FIT-502 / RO Permeate]{\label{fig:flow_exp2_p1_f}\includegraphics[scale = 0.46]{./figures/FIT_502_impact_exp1.pdf}}
\caption{Summary of Flow Over Duration of Attack to Stop Pumps in P1}
\label{fig:flow_exp2_p1}
\end{figure}

\begin{figure}[H]
\centering
\subfigure[LIT-101]{\label{fig:level_exp2_p1_a}\includegraphics[scale = 0.46]{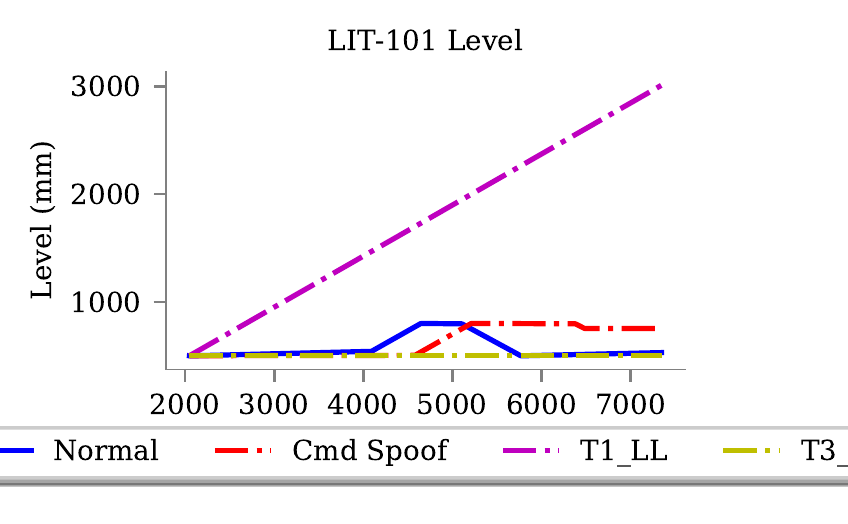}}
\subfigure[LIT-301]{\label{fig:level_exp2_p1_b}\includegraphics[scale = 0.46]{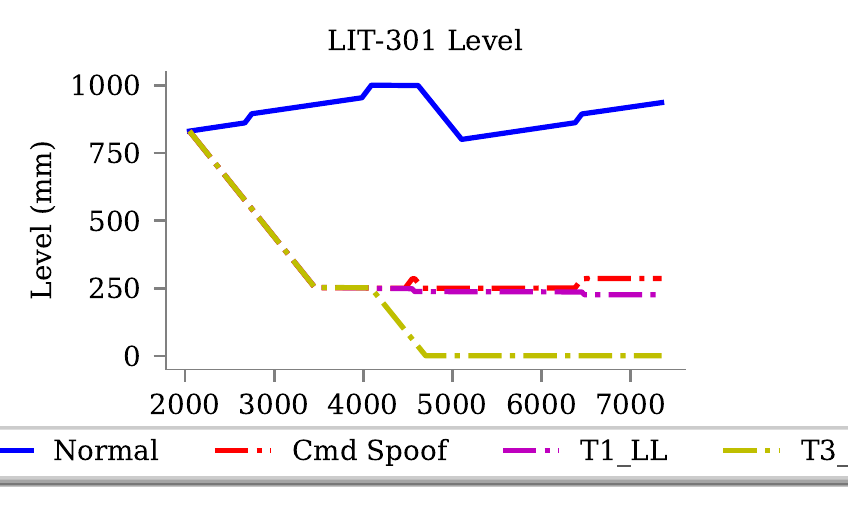}}
\subfigure[LIT-401]{\label{fig:level_exp2_p1_c}\includegraphics[scale = 0.46]{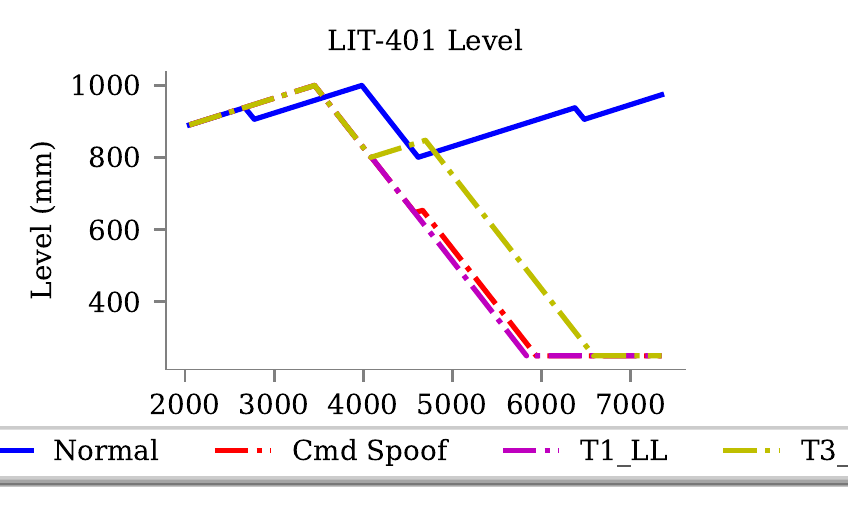}}
\caption{Summary of Tank Levels Over Duration of Attack to Stop Pumps in P1}
\label{fig:level_exp2_p1}
\end{figure}

Figures \ref{fig:flow_exp2_p1} and \ref{fig:level_exp2_p1} shows the simulation result for the flow and level sensor measurements for the experiment to stop pumps in P1.

\begin{figure}[H]
\centering
\subfigure[FIT-101]{\label{fig:flow_exp2_p3_a}\includegraphics[scale = 0.46]{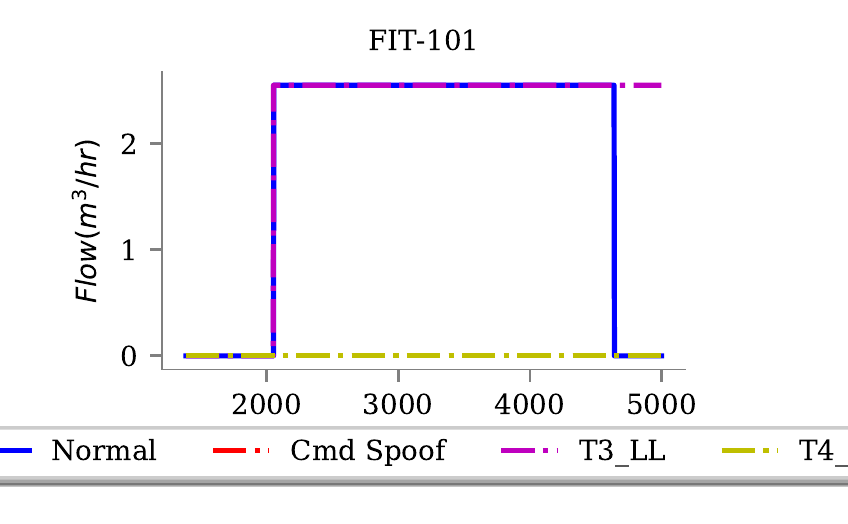}}
\subfigure[FIT-201]{\label{fig:flow_exp2_p3_b}\includegraphics[scale = 0.46]{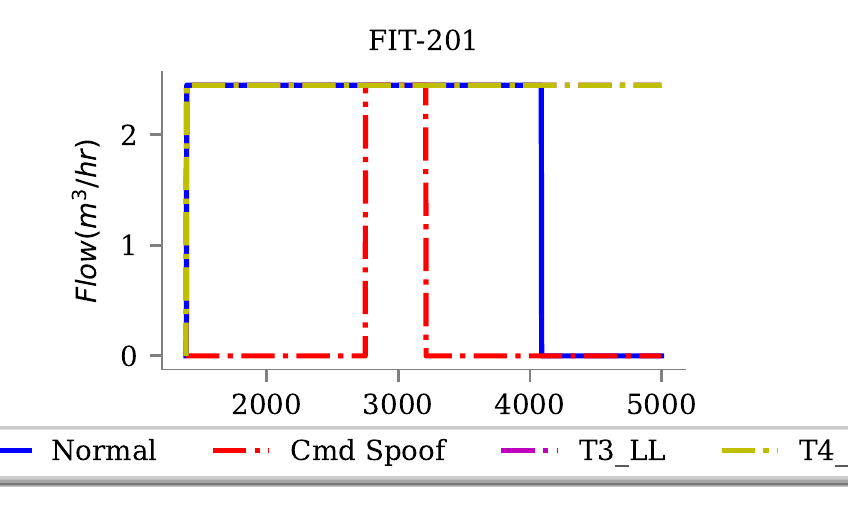}}
\subfigure[FIT-301]{\label{fig:flow_exp2_p3_c}\includegraphics[scale = 0.46]{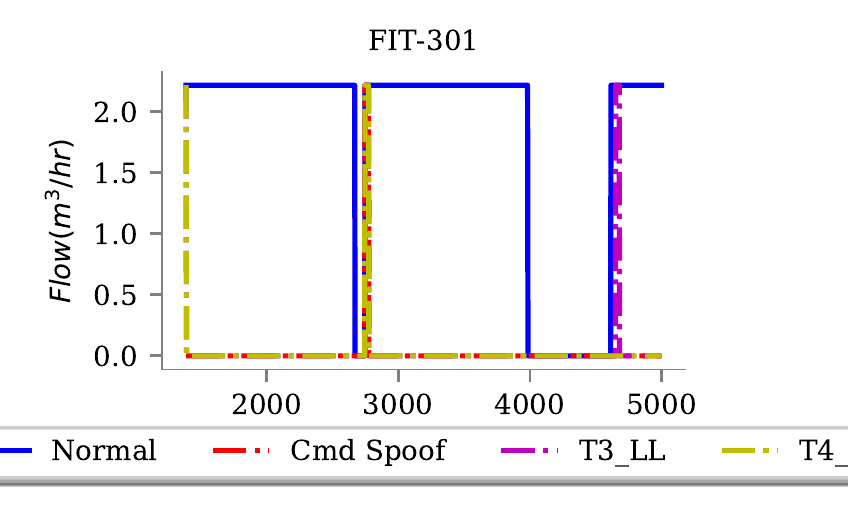}}
\subfigure[FIT-401]{\label{fig:flow_exp2_p3_d}\includegraphics[scale = 0.46]{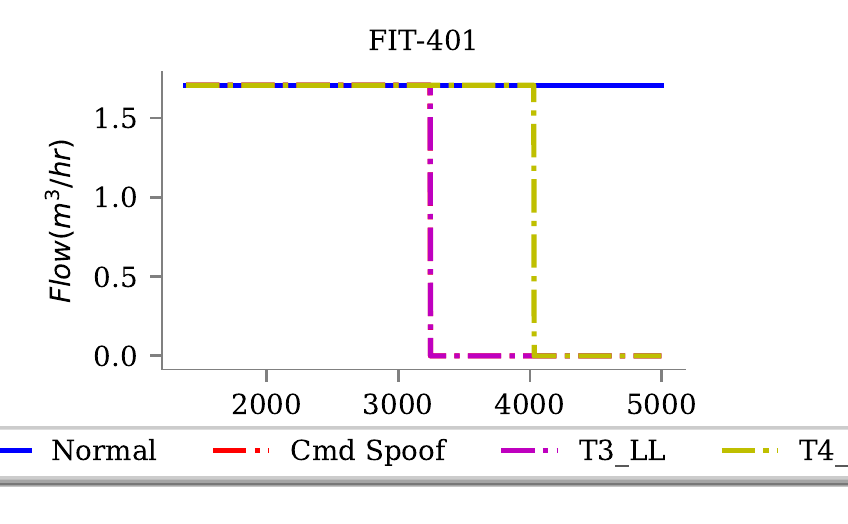}}
\subfigure[FIT-501]{\label{fig:flow_exp2_p3_e}\includegraphics[scale = 0.46]{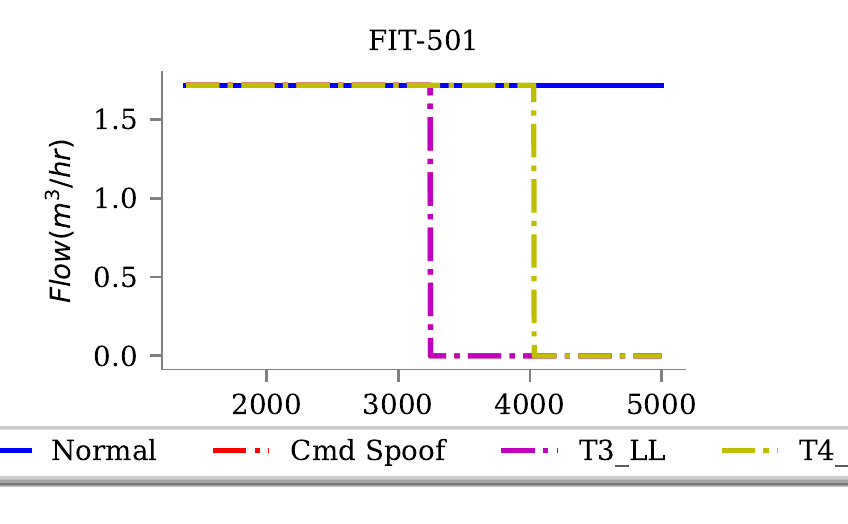}}
\subfigure[FIT-502 / RO Permeate]{\label{fig:flow_exp2_p3_f}\includegraphics[scale = 0.46]{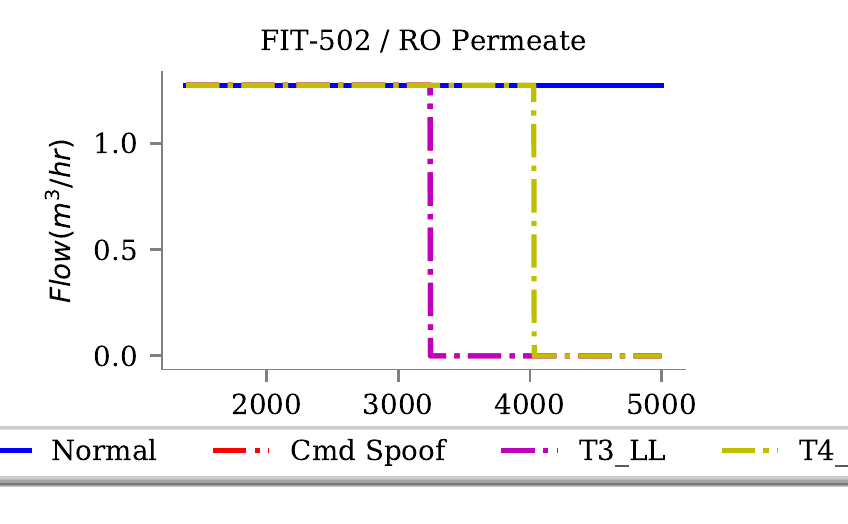}}
\caption{Summary of Flow Over Duration of Attack to Stop Pumps in P3}
\label{fig:flow_exp2_p3}
\end{figure}

\begin{figure}[H]
\centering
\subfigure[LIT-101]{\label{fig:level_exp2_p3_a}\includegraphics[scale = 0.46]{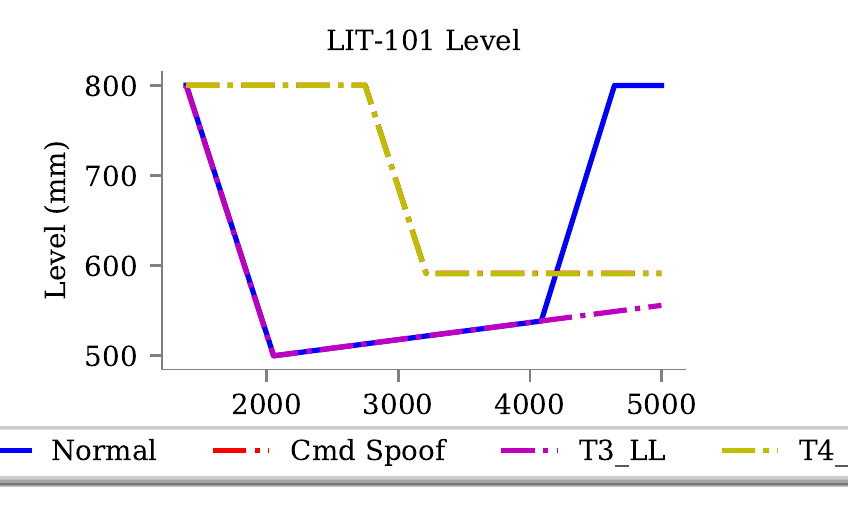}}
\subfigure[LIT-301]{\label{fig:level_exp2_p3_b}\includegraphics[scale = 0.46]{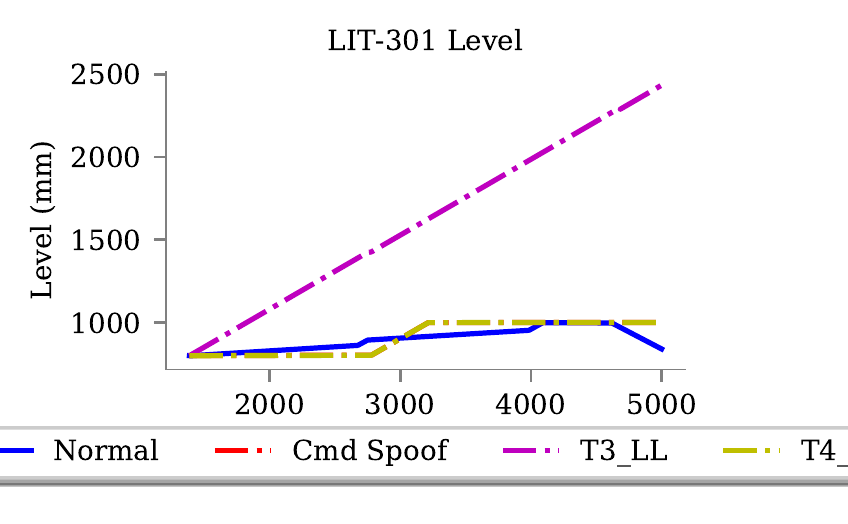}}
\subfigure[LIT-401]{\label{fig:level_exp2_p3_c}\includegraphics[scale = 0.46]{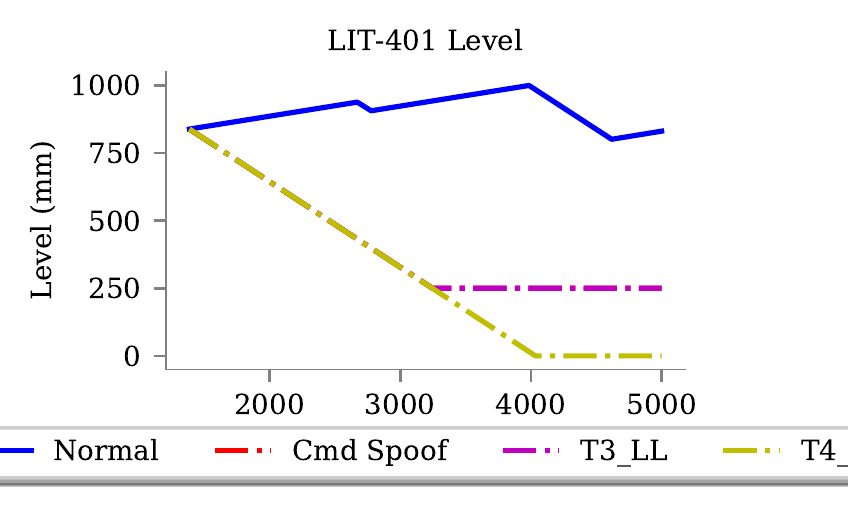}}
\caption{Summary of Tank Levels Over Duration of Attack to Stop Pumps in P3}
\label{fig:level_exp2_p3}
\end{figure}

Figures \ref{fig:flow_exp2_p3} and \ref{fig:level_exp2_p3} shows the simulation result for the flow and level sensor measurements for the experiment to stop pumps in P3.

\end{document}